\documentclass[aps, prl, 
superscriptaddress,
nobibnotes,
amsmath,
amssymb,
reprint]{revtex4-2}

\usepackage{graphicx}% Include figure files
\usepackage{dcolumn}% Align table columns on decimal point
\usepackage{booktabs}
\usepackage{xcolor, hyperref} % Put this in the preamble
\usepackage{bm, physics}% bold math
\usepackage{subfiles}  
\usepackage{mathtools}

\definecolor{DarkGreen}{RGB}{14,140,14} % A muted green

\usepackage{bibunits}
\defaultbibliography{bib} % your .bib file
\defaultbibliographystyle{apsrev4-1}
\begin{document}
\preprint{APS/123-QED}

\title{Control of Dipolar Dynamics by Geometrical Programming}

\author{Jiaqi You}
\email{jiaqiyou@g.harvard.edu}
\affiliation{Department of Physics, Harvard University, Cambridge, MA 02138, USA}
\affiliation{Harvard-MIT Center for Ultracold Atoms, Cambridge, MA 02138, USA}

\author{John M. Doyle}
\affiliation{Department of Physics, Harvard University, Cambridge, MA 02138, USA}
\affiliation{Harvard-MIT Center for Ultracold Atoms, Cambridge, MA 02138, USA}

\author{Zirui Liu}
\affiliation{Department of Physics, Harvard University, Cambridge, MA 02138, USA}
\affiliation{Harvard-MIT Center for Ultracold Atoms, Cambridge, MA 02138, USA}
\affiliation{Institute for Interdisciplinary Information Sciences, Tsinghua University, Beijing 100084, PR China}

\author{Scarlett S. Yu}
\affiliation{Department of Physics, Harvard University, Cambridge, MA 02138, USA}
\affiliation{Harvard-MIT Center for Ultracold Atoms, Cambridge, MA 02138, USA}

\author{Avikar Periwal}
\affiliation{Department of Physics, Harvard University, Cambridge, MA 02138, USA}
\affiliation{Harvard-MIT Center for Ultracold Atoms, Cambridge, MA 02138, USA}

\date{\today}
\begin{abstract}
We propose and theoretically analyze methods for quantum many-body control through geometric reshaping of molecular tweezer arrays. Dynamic rearrangement during entanglement is readily available due to the extended coherence times of molecular rotational qubits. We show how motional dephasing can be suppressed and enhanced spin squeezing can be achieved in an actively rearranged short-range XY model. We also analyze in detail a specific static geometry that significantly suppresses decoherence. These general methods as applied to programmable quantum systems offer robust control modalities that are well suited to molecules.
\end{abstract}

\maketitle
\begin{bibunit} % Main references
Ultracold polar molecules in optical tweezer arrays are a powerful platform for quantum simulation, combining long-lived rotational states~\cite{burchesky2021rotational, park2023extended, gregory2024second}, long-range dipolar interactions~\cite{ni2018dipolar, holland2023demand, picard2025entanglement}, rich internal structure~\cite{hepworth2024coherent}, and manipulation of individual particles with high spatial control~\cite{anderegg2019optical, vilas2024optical}. A key feature is the exceptionally long coherence time of rotational qubit states, exceeding several seconds~\cite{park2017second, ruttley2025long}, which is orders of magnitude longer than the timescales required for internal state control~\cite{silveri2017quantum, blackmore2020controlling} and spatial rearrangement~\cite{barredo2016atom, endres2016atom, bluvstein2022quantum}.  These features present a regime where extensive classical resources—computation, optimization, and spatial manipulation—can be deployed in real time during coherent many-body evolution. In this context, array geometry can potentially become a time-dependent control knob offering direct and flexible means of engineering interactions.  In optical tweezer arrays, arbitrary, controlled placement and movement of individual particles can be performed with sub-wavelength precision~\cite{barredo2016atom, endres2016atom, deist2022superresolution}. Spatial configuration and orientation of particles specify the strength and sign of interactions, and can thus control quantum dynamics. Envisioned possibilities that take advantage of these properties include real-time geometric reshaping, which offers both Hamiltonian engineering and non-equilibrium quantum control.

In this Letter, we present theoretical studies of how programmable control over geometry, both static and dynamic, gives rise to useful capabilities in dipolar quantum systems. In the static setting, engineering fixed geometries can be used to suppress thermally induced decoherence of dipolar interactions. In the dynamic setting, the geometry can be actively reshaped during coherent entanglement evolution, enabling both corrective and constructive applications. Dynamically rearranging molecular positions can be controlled to cancel phase shifts induced by thermal motion, realizing robust dipolar interactions in one- and two-dimensional arrays. This same capability allows for programmable control over many-body spin dynamics. We theoretically demonstrate that rearranging molecules in a tweezer array during qubit phase evolution can enhance spin squeezing generated by dipolar dynamics beyond what is possible in static configurations. These results establish geometry-controlled dipolar dynamics as a dynamic analog paradigm, complementary to both static analog simulation based on fixed geometries~\cite{davis2023probing, browaeys2020many, bornet2023scalable, su2023dipolar}, and gate-based digital approaches~\cite{raeisi2012quantum, bluvstein2022quantum, bluvstein2024logical}.

We present three findings: static geometrical suppression of thermally induced dephasing; programmable geometrical control creating robust multiparticle entanglement; and shaping of many-body quantum dynamics for metrological squeezing. The physics of the first finding lays the foundational understanding for the other findings. These examples highlight the power of geometric control for dipolar Hamiltonian engineering.

%We begin with the static example by addressing a key challenge: decoherence arising from thermal motion.
With our first finding, we study in detail a choice of static geometry that suppresses thermally induced dephasing.  This builds on the physics described in ref.~\cite{ni2018dipolar}. Although polar molecules offer intrinsically long-lived rotational coherence---with timescales far exceeding the timescale of dipolar interactions---achieving high-fidelity entanglement remains challenging in practice. A primary limitation is thermal motion~\cite{bao2023dipolar, holland2023demand, ruttley2025long, picard2025entanglement, bergonzoni2025iswap}, which contributes significantly to decoherence in both directly laser-cooled and assembled molecule platforms. For laser-cooled species such as CaF, gray molasses cooling~\cite{anderegg2018laser, cheuk2018lambda} in the tweezer typically results in thermal motion well outside the Lamb-Dicke regime. Raman sideband cooling can reduce motional excitation to near the ground state, but the procedure is experimentally demanding and often incurs a substantial loss of molecules, primarily due to limited optical pumping efficiency in molecular systems~\cite{lu2024raman, caldwell2020sideband, bao2024raman}. For Feshbach-assembled molecules such as NaCs or RbCs, residual motional fluctuations also introduce appreciable dephasing in dipolar spin dynamics~\cite{ruttley2025long, picard2025entanglement}.

Thermal dephasing can be understood semi-classically in the regime where dipolar interactions are much slower than trap frequencies~\cite{bao2023dipolar, holland2023demand, ruttley2025long, picard2025entanglement}. In this limit, fast oscillatory motion averages out short-time fluctuations. However, the oscillation amplitude, sampled from a thermal distribution, changes the time-averaged dipolar interaction strength, resulting in interaction strength disorder between experimental realizations~\cite{SM}. For a thermal distribution in a harmonic trap, the standard deviation of the oscillation amplitude is approximately 50\% of its mean. This leads to significant dephasing of the accumulated dipolar interaction phase between molecules.

\begin{figure}[t] \centering\includegraphics[width=0.9\columnwidth]{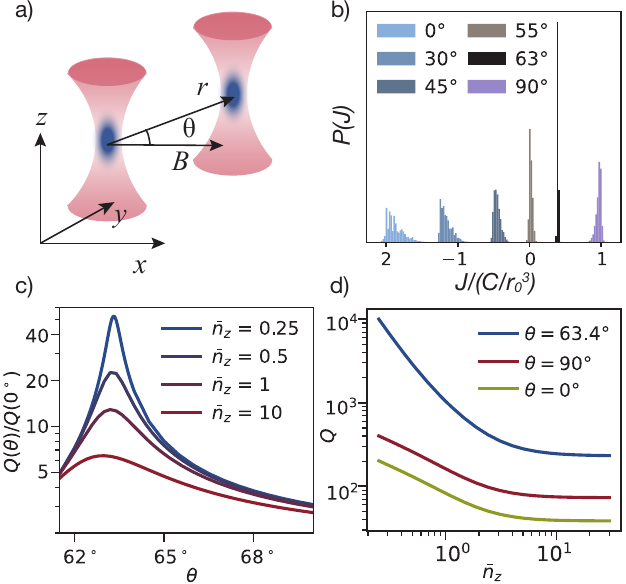}
    \caption{Angular dependence of dipolar interactions in the presence of thermal motion.
    (a) Molecules trapped in optical tweezers (red) have thermally broadened spatial distributions (blue). The dipole alignment is set by the external magnetic field $\mathbf{B}$, forming an angle $\theta$ with the interparticle displacement vector $\vec{r}$ in the $x$-$y$ plane.
    (b) Distribution of dipolar couplings $J$ for various field angles $\theta$, showing minimal width at the magic angle, where leading-order sensitivity to axial motion vanishes.
    (c) Quality factor $Q = \tau |J|$ (normalized to its value at $0^\circ$) versus $\theta$ for different averaged axial thermal occupations $\bar{n}_z$. (d) Quality factor $Q$ versus $\bar{n}_z$ for different $\theta$. Smaller thermal extent leads to stronger suppression of interaction noise and a sharper optimum near the magic angle. We assume a thermal distribution over motional states \( \{n_x, n_y, n_z\} \), corresponding to the same temperature along all three directions. Simulations are performed for CaF molecules trapped in a tweezer array with axial (radial) trap frequencies of 20\,kHz (100\,kHz) and an interparticle spacing of $r = 2\,\mu\mathrm{m}$. See the supplementary material~\cite{SM} for simulation details.}
    \label{fig:magic_angle}
\end{figure}

% In our first finding, we show that a suitable choice of static geometry can suppress thermally induced dephasing.  This builds on the physics described in ref~\cite{ni2018dipolar}.
Intrinsically, the interaction strength $J$ between two particles varies with both the angle and the distance between them. Thus, position fluctuations due to thermal motion lead to geometry-dependent fluctuations in $J$. The geometry can be appropriately chosen to mitigate motion-induced dephasing by suppressing the sensitivity of $J$ to thermal fluctuations. We analyze a representative configuration as shown in Fig.~\ref{fig:magic_angle}(a). The quantization axis is along the $\hat{x}$ direction and the molecular pair lies along a separation vector \( \vec{r} \), with small relative displacements \( \delta\vec{r} = (x, y, z) \) between two molecules in the center-of-mass frame. The dipolar interaction is given by 
\begin{equation} 
J(\vec{r}+\delta\vec{r} ) = \frac{C}{|\vec{r} + \delta\vec{r}|^3} \left(1 - 3\cos^2\theta(\vec{r}+\delta\vec{r}) \right), \end{equation} 
where $\theta$ is the angle between the dipole orientation and the interparticle vector. For $x, y, z \ll r$, we expand $J$ to leading orders: 
\begin{align}
J(\vec{r} + \delta \vec{r}) \approx \frac{C}{r^3} \bigg[
& 1 - 3\cos^2\theta(\vec{r}) \notag + A_{y,1}(\theta(\vec{r})) \frac{y}{r} \\
& + A_{z,2}(\theta(\vec{r})) \frac{z^2}{r^2} + A_{x,2}(\theta(\vec{r})) \frac{x^2}{r^2} + \cdots \bigg].
\label{eq:expansion}
\end{align}
Here, the expansion coefficients \( A_{i,j} \) are dimensionless geometric factors that depend on the angle \( \theta(\vec{r}) \), corresponding to motion along the $\hat{i}$ axis. In the regime where the trap frequency greatly exceeds the dipolar interaction rate, fast thermal motion averages out linear terms such as $A_{y,1}y/r$, and the dominant contribution to the variation of \( J \) arises from second-order terms $A_{i, 2}$. Moreover, in typical optical tweezers, the confinement is highly anisotropic, as the cloud width along the axial direction is much larger than in the radial plane, making axial motion the dominant source of thermally induced dephasing.

To leading order, the disorder of the interaction strength $J$ due to thermal motion can be expressed as $\eta_{z,2} = A_{z,2}\langle z^2 \rangle/r^2$. Here, \( \langle z^2 \rangle \) denotes the mean squared axial displacement within a single experimental realization. While $\langle z^2 \rangle$ is sampled from a thermal distribution, $A_{z,2}/r^2$ can be tuned---in both sign and magnitude---by modifying the geometry of the dipolar interaction. This tunability is illustrated in Fig.~\ref{fig:magic_angle}(b), which shows the distribution of \( J \) values at various dipole angles \( \theta \). In particular, at a ``magic angle'' $\theta_m = 63.4^\circ$, we have $A_{z,2}=\frac{3}{4}(3 + 5\sin (2\theta_m)) =0$, making $J$ insensitive to axial thermal motion at the leading order. Crucially, at $\theta_m$, the mean interaction remains non-zero, with \( J \propto 1 - 3\cos^2\theta_m = 0.4 \).  As $\theta \rightarrow \theta_m$ in Fig.~\ref{fig:magic_angle}(b), the width of the \( J \) distribution narrows significantly, indicating suppression of thermal disorder. 
% This magic condition can be viewed as controlling the geometry of interacting dipoles, or as controlling the external field that is aligning them.  
% A similar magic condition has been found for experiments where the quantization axis is primarily aligned along the tweezer $k$-vector~\cite{ni2018dipolar}.
% This magic-angle configuration enables robust dipolar coupling in the presence of motion, without requiring cooling to the motional ground state. 

One limitation of this magic-angle condition is that it does not cancel out higher order sensitivity to motion beyond the $\langle z^2 \rangle$ term. The effectiveness of the $A_{z,2}$ decoupling depends on two geometric ratios. First, the decoupling improves as the relative spacing of the two traps with respect to the thermal extent of each particle increases, characterized by $A_{z, 2}/\ev{\eta_{z, 2}}$, where the expectation values denote the average over the entire thermal distribution. As shown in Fig.~\ref{fig:magic_angle}(c)-(d),  lower axial thermal occupation $\bar{n}_z$ leads to a stronger suppression of interaction-induced dephasing near the magic angle, as quantified by the quality factor $Q=\tau |J|$, where $\tau$ is the damping time for spin exchange oscillations~\cite{emperauger2025benchmarking}. Second, suppression of thermal motion improves as the anisotropy of the trapping frequencies \( \alpha = A_{x,2}\langle \eta_{z, 2} \rangle / A_{z, 2}\langle \eta_{x, 2}\rangle = \omega_{\text{radial}}^2 / \omega_{\text{axial}}^2 \) increases. Larger values of $\alpha$ lead to a stronger fractional suppression of the overall thermal motion. Residual decoherence remains in all cases due to higher-order axial contributions (e.g., \( \langle z^4 \rangle \)) and due to radial motion along \( x \) and \( y \),  whose contributions cannot be simultaneously eliminated in a single static configuration. In addition to these residual couplings, the magic-angle condition applies only to repulsive configurations and to specific dipole-field alignments. As a result, not all pairs of molecules in higher-dimensional arrays can simultaneously satisfy this condition, limiting its applicability to simulate many-body systems. 

\begin{figure}[t]
    \centering
    \includegraphics[width=0.9\columnwidth]{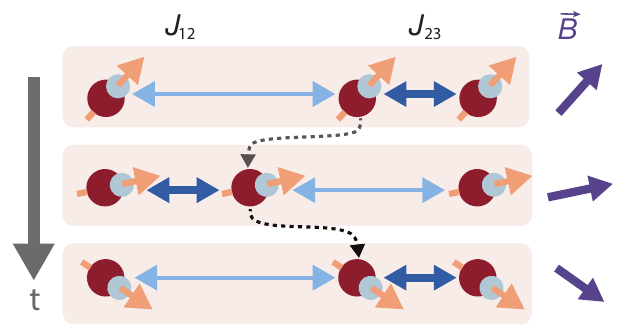}
    \caption{Dynamical control of dipolar Hamiltonians via geometrical programming. Time-dependent control over both molecular positions (black dashed arrows) and dipole orientations via the external field $\hat{\mathbf{B}}(t)$ (purple arrows) enables reconfiguration of the dipolar interaction graph, at timescales much faster than both dipolar interaction and coherence times. This capability allows dynamic tuning of the anisotropic dipolar couplings, $J_{ij}(t) \propto \left(1 - 3\cos^2\theta_{ij}(t)\right)/r_{ij}^3(t)$ (blue bonds), during coherent many-body evolution.}
    \label{fig:platform}
\end{figure}

In our second finding, we show how real-time dynamic control of system geometry enables more robust and flexible protection against motional dephasing. We devise a ``geometric echo'' protocol as a simple prototype for dynamic geometrical Hamiltonian tuning. The essential mode of operation is shown in Fig.~\ref{fig:platform}. In this dynamic framework, the system evolves under a sequence of discrete geometries $\{ \vec{r}_i \}$ for durations $t_i$, such that motion-induced phase shifts cancel in the ensemble average.  We assume that the particles' motional states remain fixed throughout the sequence, so that dephasing can be coherently canceled across geometric segments. For example, the leading-order axial dephasing term between molecules $a$ and $b$ can be eliminated if the weighted sum of geometric sensitivities vanishes:
\begin{align}
\sum_i \frac{A_{z,2}(\theta_{ab,i}) t_i}{|\vec{r}_{a,i} - \vec{r}_{b,i}|^2} = 0.
\label{eq:axial_sensitivity}
\end{align}
This approach supports more sophisticated control strategies. For instance, one can construct sequences that preserve a target nonzero effective interaction \( J_{\mathrm{eff}} = \sum_i J_i t_i / \sum_i t_i \), simultaneously decouple a molecule from multiple dipolar neighbors in higher-dimensional arrays, or suppress higher-order contributions such as those proportional to the $\ev{z^4}$ or $\ev{x^2}$ terms.

\begin{figure}[t]
    \centering\includegraphics[width=0.9\columnwidth]{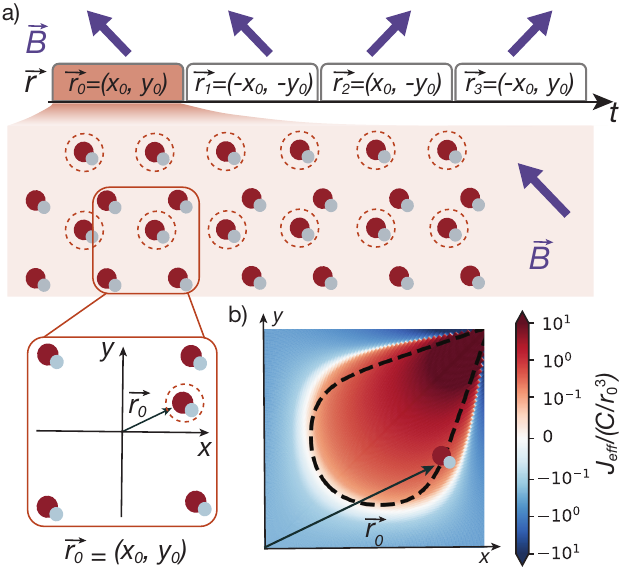}
    \caption{Suppression of leading-order thermal dephasing via geometry echo in a square array.
    (a) Schematic of a four-step ``bang-bang'' protocol that cyclically reconfigures the geometry of a square array by rearranging half of the molecules (marked with dashed circles) with relative displacement $\vec{r}_i$. The displacement vectors \(\vec{r}_0, \vec{r}_1, \vec{r}_2, \vec{r}_3\) satisfy symmetry constraints (as illustrated) to preserve the spatial symmetry of the square array over a full cycle. Dipole orientations are rotated at each step via control of the external magnetic field $\vec{B}$. 
    (b) Effective dipolar interaction $J_\mathrm{eff}$ (averaged over a full echo cycle) as a function of relative displacement $\vec{r}_0=(x_0, y_0)$ used in the first step. The dashed black contour indicates configurations for which the sum of geometric sensitivities, as defined in Eq.~\eqref{eq:axial_sensitivity}, from all four nearest neighbors vanishes, thereby canceling leading-order dephasing. Selecting a representative point $\vec{r}_0$ on this contour and applying the protocol in (a) enables robust decoupling of thermal motion from nearest-neighbor dipolar interactions in square arrays while maintaining a non-zero effective interaction strength \( J_{\mathrm{eff}} \).}
    \label{fig:echo}
\end{figure}

As a concrete example, we demonstrate a protocol that can be implemented in a two-dimensional square array to realize XX-model dynamics with suppression of leading-order thermal dephasing, through a sequence of non-trivial rapid jumps in positions. As illustrated in Fig.~\ref{fig:echo}(a), the main idea is to cycle through four geometries by rearranging half of the molecules with displacements $\vec{r}=\{\vec{r}_0, \vec{r}_1, \vec{r}_2, \vec{r}_3\}$ while coherently rotating the dipole orientation via external field control. In each step, half of the molecules remain in static tweezers while the other half are dynamically repositioned in parallel, enabling scalable implementation across large arrays.

To evaluate the effectiveness of this multi-step echo protocol, we compute the geometry-averaged axial sensitivity, as defined in Eq.~\eqref{eq:axial_sensitivity}, and the effective interaction strength $J_{\mathrm{eff}}$ accumulated across the four steps as a function of relative displacement $\vec{r_0}=(x_0, y_0)$ used in the first step, shown in Fig.~\ref{fig:echo}(b). While the color shows the effective interaction strength, the dashed line shows the contour on which motional dephasing from all four nearest neighbors is canceled. Owing to the preserved spatial symmetry of the four-step protocol, the decoupling condition satisfied for one neighbor of a central molecule automatically applies to the other three neighbors. Importantly, the dephasing-cancellation contour intersects regions with finite \( J_{\mathrm{eff}} \), enabling robust dipolar interactions. This scalable protocol illustrates how dynamic control of geometry can overcome the limitations of static configurations and enable flexible and robust many-body evolution.

The geometric echo method as described above is just one example of what might be accomplished. Looking forward, geometric echo protocols could be further optimized to achieve more ambitious objectives, such as encoding a target \( J_{\mathrm{eff}} \) for an arbitrary pair of molecules, suppressing higher-order thermal effects, or decoupling from multiple neighbors with different dipolar angles. In a laboratory setting these protocols will also require robustness to technical imperfections, such as imperfect alignment between different tweezer sites. 

We also note that in some cases, continuously varying the geometry, rather than stepping through discrete configurations, could provide more stable control or improved cancellation. Advances made possible in this way would further enhance the practicality and flexibility of geometry as a programmable tool for dipolar Hamiltonian engineering.

Our third finding uses more complex programmable geometrical control over ensembles of molecules to actively shape many-body dipolar spin dynamics. This requires classical control of particle positions on timescales much faster than the interactions. For demonstrated interaction strengths in molecular tweezer arrays and rearrangement speed of particles with optical tweezers, the difference in the interaction timescale when compared to the rearrangement timescale can be more than a factor of 100~\cite{SM}. In our illustrative example, we show that dynamic rearrangements can generate scalable spin squeezing and yield near-maximal metrological gain.

\begin{figure}[t]
    \centering    \includegraphics[width=\columnwidth]{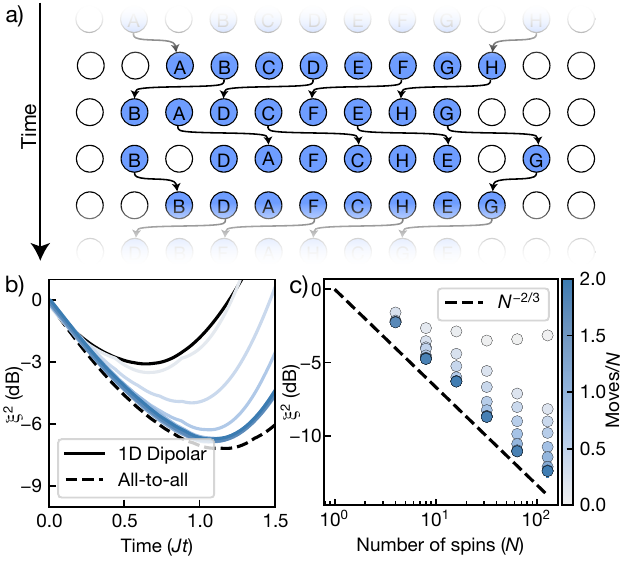}
    \caption{Enhanced spin squeezing with dynamic rearrangement 
    (a) Illustration of a rearrangement protocol.
    (b) Squeezing parameter $\xi^2$ calculated from DTWA plotted vs. time for $N = 20$ spins.  Colors indicate an increasing number of repetitions of the move shown in (a). (c) Optimal squeezing for different system sizes and ``move'' steps.  The dashed black line shows the expected scaling for all-to-all interactions. The colored points show DTWA simulations, with darker blue circles show increasing number of moves.}
    \label{fig:squeezing}
\end{figure}

Dipolar interactions in ordered arrays of quantum particles naturally produce spin-squeezed states~\cite{perlin2020spin, bornet2023scalable, eckner2023realizing, hines2023spin, block2024scalable, douglas2024spin}.  However, the relative speed of generating entanglement is fundamentally limited by the range of interactions. In the ideal regime to maximize squeezing, the interaction strength is uniform across all spin pairs, preserving permutation symmetry. This symmetry ensures that squeezing grows with system size, achieving a scaling of $\xi^2 \sim N^{-2/3}$, where $\xi^2$ is the Wineland squeezing parameter~\cite{wineland1994squeezed} and $N$ is the number of particles.

The spatially localized couplings of dipolar arrays break permutation symmetry, and the resulting dynamics deviate from the ideal all-to-all case.  As the number of neighbors of each dipole increases, the maximum achievable metrological gain improves~\cite{wellnitz2024spin}. In one-dimension, squeezing is not scalable, and the achievable gain saturates~\cite{block2024scalable}. By dynamically reconfiguring the spatial geometry of the array during evolution, we can stitch together different interaction graphs to simulate an effectively more uniform interaction landscape. For example, by evolving between pairwise interactions, each spin can interact with multiple partners over time, approximating the collective effect of global interactions.

In Fig.~\ref{fig:squeezing}(a), we illustrate a set of three simple moves, which in aggregate permute a 1D array. This permutation on $N$ elements shifts any individual molecule's neighbors by two, and rotates the two boundary sites. Effectively, each molecule interacts along a conveyor belt formed by the other molecules. If repeated $N$ times, this procedure causes each molecule to interact for equal time with every other molecule, up to small boundary effects. In the many-move limit, the average Hamiltonian has an all-to-all nature.  This effect can be seen in Fig.~\ref{fig:squeezing}(b), where we plot achievable squeezing for 20 spins in a one-dimensional array as a function of time, expressed in units of $J$, calculated using the Discrete Truncated Wigner Approximation (DTWA)~\cite{schachenmayer2015many}. Adding more permutations (darker colors) increases the achievable metrological gain, which saturates at the all-to-all bound, shown as dashed black line.  We then examine the scaling of the resultant squeezing as a function of the number of spins $N$, shown in Fig.~\ref{fig:squeezing}(c).  The dashed black line shows the expected scaling for all-to-all interactions.  The colored points show DTWA simulations, where darker colors represent an increasing number of permutations.  For a small number of permutations, the effective Hamiltonian is quasi-1D, and does not exhibit scalable spin-squeezing.  However, with $O(N)$ moves, we recover the optimal all-to-all scaling.

In this way, dynamic rearrangement during entanglement allows us to recover all-to-all squeezing dynamics, even in systems with fundamentally short-range interactions. The effective interaction becomes more homogeneous, enabling long-range entanglement in shorter times, and thus improving the accessible limits of spin squeezing with spatially local interactions. This approach opens avenues for programmable spin dynamics and metrological enhancement in dipolar molecular arrays.

In summary, we demonstrate that geometrical control enables useful improvements and new directions in dipolar quantum systems. Dynamic geometrical tuning is shown to be a versatile control modality for robust and programmable quantum systems, a paradigm that is perfectly suited to molecular tweezer arrays. Some of these approaches to geometric control during interactions could also be applied to other systems with tunable strength interactions, including Rydberg-dressed atoms~\cite{borish2020transverse, schine2022long}. For molecular systems, more complicated protocols could employ large amounts of classical computational power during the entanglement time. This may open new avenues to not only quantum simulation but also robust quantum computation~\cite{ma2023high, scholl2023erasure}. 

\section*{Acknowledgments}
We thank Lo\"ic Anderegg, Eunmi Chae, Youngju Cho, Wolfgang Ketterle, Qinshu Lyu, Kang-Kuen Ni, and Norman Yao for valuable discussions. This material is based upon work supported by the U.S. Department of Energy, Office of Science, National Quantum Information Science Research Centers, Quantum Systems Accelerator. Additional support is acknowledged from Harvard-MIT Center for Ultracold Atoms (Grant No. PHY-2317134); the Air Force Office of Scientific Research (AFOSR)'s AOARD under award number FA2386-24-1-4070; and from MURI W911NF-19-1-0283.

\putbib
\end{bibunit}
\clearpage
\onecolumngrid
\begin{center}
\textbf{\large Supplementary Material}
\end{center}
\setcounter{section}{0}
\setcounter{figure}{0}
\setcounter{table}{0}
\renewcommand{\thesection}{S\arabic{section}}
\renewcommand{\thefigure}{S\arabic{figure}}
\renewcommand{\thetable}{S\arabic{table}}
\begin{bibunit} % Supplement references

\ifSubfilesClassLoaded{%
  \title{Supplementary Material}
  \date{\today}
  \maketitle
}{}

\onecolumngrid

In this supplementary material, we discuss the mathematical modeling used to calculate thermally induced dephasing at different temperatures and angles. We also discuss reasonable timescales based on experimental parameters for several polar molecular species in existing tweezer and lattice experiments. Finally, we provide two different rearrangement methods for enhancing spin-squeezing in arrays of dipolar molecules. 

\section{Dipolar Interaction Strength}
We evaluate the strength and motional sensitivity of the dipolar spin-exchange interaction between a pair of molecules. The interaction arises from the electric dipole-dipole Hamiltonian, which can be written as
\begin{equation}
\hat{H}_{\mathrm{dd}} = \frac{1}{4\pi\varepsilon_0} \frac{1}{r^3} \left( \mathbf{d}_i \cdot \mathbf{d}_j - 3 (\mathbf{d}_i \cdot \hat{\mathbf{r}})(\mathbf{d}_j \cdot \hat{\mathbf{r}}) \right),
\end{equation}
where $\mathbf{d}_i$ and $\mathbf{d}_j$ are the dipole operators of molecules $i$ and $j$, $\hat{\mathbf{r}} = \mathbf{r}/|\mathbf{r}|$ is the unit vector pointing from one dipole to the other, and $r = |\mathbf{r}_i - \mathbf{r}_j|$. In spherical tensor notation, the Hamiltonian can be recast as:
\begin{equation}
\hat{H}_{\mathrm{dd}} = -\frac{1}{4\pi\varepsilon_0} \sqrt{6} \, T^2_p(\mathbf{C}) \cdot T^2_p(\mathbf{d}_i, \mathbf{d}_j).
\end{equation}
Here $T_p^2(\mathbf{C})=\langle | C_p^2(\theta, \phi) / r^3| \rangle$ is the $p^{th}$ component of a rank-2 tensor, which encodes the geometry-dependent coupling  in terms of the spherical harmonic $C^2_p(\theta, \phi)$. The second term, $
T_p^2\left(\mathbf{d}_i, \mathbf{d}_j\right)=T^1\left(\mathbf{d}_i\right) \times T^1\left(\mathbf{d}_j\right)$, is the tensor product of the rank-1 spherical tensor representation of dipole operators. In this formulation, thermal motion introduces fluctuations in $T^2(\mathbf{C})$. By designing geometries in which $T^2(\mathbf{C})$ is less sensitive to positional perturbations, one can mitigate dephasing during the dipolar exchange interaction.

We focus here on the on-resonance spin-exchange interactions in the XY model, corresponding to the $p=0$ component:
\begin{equation}
\hat{H}_{\mathrm{dd}} = -\frac{1}{4\pi\varepsilon_0} \sqrt{6} \, T_0^2(\mathbf{C}) \, T_0^2(\mathbf{d}_i, \mathbf{d}_j)
= \frac{d^2}{4\pi\varepsilon_0}(1 - 3\cos^2\theta) (\left|\uparrow\downarrow\right\rangle \left\langle\downarrow\uparrow\right| + \mathrm{h.c.}),
\end{equation}
where $d$ is the transition dipole moment between $\left|\downarrow\right\rangle$ and $\left|\uparrow\right\rangle$, and $\theta$ is the angle between the quantization axis and the interparticle vector $\mathbf{r}$. In this form, we introduce the spin-exchange oscillation frequency $J$, so that: 
\begin{equation}
\hat{H}_{\mathrm{dd}} = \frac{J}{2} \left( \left|\uparrow\downarrow\right\rangle \left\langle\downarrow\uparrow\right| + \left|\downarrow\uparrow\right\rangle \left\langle\uparrow\downarrow\right| \right).
\end{equation}

In general, different magic conditions exist for different components of the full dipole-dipole Hamiltonian, and could be explored under specific choices of external fields. To assess the influence of thermal motion on the interaction ($J$), we employ two complementary models, one classical and one quantum, which we introduce in the following two sections.

\subsection{Effects of motional excitation: classical approach}

For sufficiently warm molecules, their motion may be treated semi-classically. In this model, we expand the dipolar Hamiltonian in powers of position excursions around the trap centers. This allows analytic expressions for how each coordinate contributes to disorder in the interaction, offering intuitive guidance for geometric engineering. Specifically, we write:
\begin{equation}
\frac{(1 - 3\cos^2\theta(\vec{r} + \delta\vec{r})) }{ |\vec{r} + \delta\vec{r}|^3} = \frac{1}{r^3}\bigg[
 1 - 3\cos^2\theta(\vec{r}) + A_{y,1}(\theta(\vec{r})) \frac{y}{r}
 + A_{z,2}(\theta(\vec{r})) \frac{z^2}{r^2} + A_{x,2}(\theta(\vec{r})) \frac{x^2}{r^2} + \cdots \bigg]
\end{equation}

In our system, the trapping potential is essentially harmonic and approximately symmetric about the trap center. In the regime where the trap frequency greatly exceeds the dipolar interaction rate, fast thermal motion averages out linear fluctuations.  Thermal motion thus leads to symmetric position distributions, for which odd-order moments vanish. Therefore, we neglect odd-order (antisymmetric) terms in the expansion and retain only even-order (symmetric) terms to describe motional dephasing.

The dominant contributions to disorder in the interaction arise from second-order fluctuations. The most relevant terms are listed in Table~\ref{tab:sensitivity}. These coefficients are additionally plotted in Fig.~\ref{fig:semiclassical_coefficients} as a function of the dipole orientation angle $\theta$.  Notably, the zero-crossings for $A_{x, 2}, A_{y, 2}, A_{z, 2},$ and $A_{z, 4}$, corresponding to points of thermal insensitivity, all occur at dipole orientation angles that retain a non-zero interaction strength $J$. In fact, at the angle of $\theta_m \approx 63.4^\circ$, where axial sensitivity $A_{z, 2}$ is suppressed, we see additional suppression of the radial components $A_{x, 2}$ and $A_{y, 2}$.

% [PUT IN POINTER TO KKN PAPER, REF 4 and explain that they considered the static case only, here we give a detailed analysis, but the general conclusion is the same for static case (or something like that)]

We note that a related analysis was carried out in Ref.~\cite{ni2018dipolar}, considering only the static case. There, the quantization axis was aligned with the tweezer axial axis, and thermal insensitivity requires tuning of the external magnetic field to a magic angle, or moving tweezers out of the focal plane, which is technically challenging. In our work, with the field perpendicular to the tweezer axis, $\theta$ can be tuned by rearranging molecule positions in a two-dimensional focal plane, or through external field control.

% While the general conclusion—that certain orientations suppress leading thermal dephasing—matches the static case, we provide a detailed coefficient analysis for the dynamic geometry regime enabled by mid-interaction rearrangement.

\begin{table}[h]
\centering
\begin{tabular}{lc}
\toprule
\textbf{Term} & \textbf{Sensitivity Coefficient} \\
\midrule
$x^2$ & $A_{x,2} =-3/16 \left( 9 + 20 \cos 2\theta + 35 \cos 4\theta \right)$ \\
$y^2$ & $A_{y,2} =3/16 \left( -3 + 35 \cos 4\theta \right)$ \\
$z^2$ & $A_{z,2} =3/4\left( 3 + 5 \cos 2\theta \right)$ \\
$z^4$ & $A_{z,4} = 15/16 \left(5 + 7 \cos 2\theta \right)$\\
% $x^2 z^2$ & $A_{x,2,z,2} =32/45 \left( 15 + 28 \cos 2 \theta + 21 \cos 4\theta \right)$ \\
% $y^2 z^2$ & $A_{y,2,z,2}=32/45 \left( 5-21 \cos 4\theta \right)$ \\
\bottomrule
\end{tabular}

\caption{Sensitivity coefficients $A_{i,n}$ characterizing the coupling between thermal disorder and dipolar interaction strength $J$. The coefficients are derived from a Taylor expansion of $J$ around the mean trap center, retaining only even-order terms.}
\label{tab:sensitivity}
\end{table}

\begin{figure}[h!]
\centering
\includegraphics[width=0.6
\textwidth]{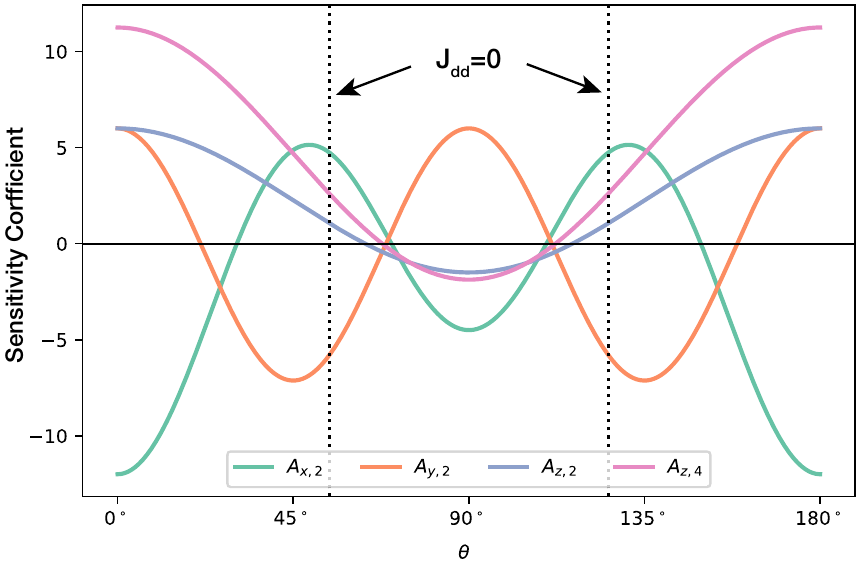}
\caption{Angular dependence of the sensitivity coefficients $A_{i,n}$, which quantify how thermal motion along the $i$ axis contributes to fluctuations in the dipolar coupling strength $J$. Notably, $A_{z,2}$ vanishes near $\theta \approx 63.4^\circ$, indicating suppressed sensitivity to motion along $\hat{z}$ at that angle. This is distinct from $\theta \approx 54.7^\circ$, where the dipolar interaction $J_{dd} \propto 1 - 3\cos^2 \theta$ vanishes, indicated by the dashed vertical lines.}
\label{fig:semiclassical_coefficients}
\end{figure}

\subsection{Effects of motional excitation: quantum approach}

\begin{figure}[h!]
\centering
\includegraphics[width=0.95\textwidth]{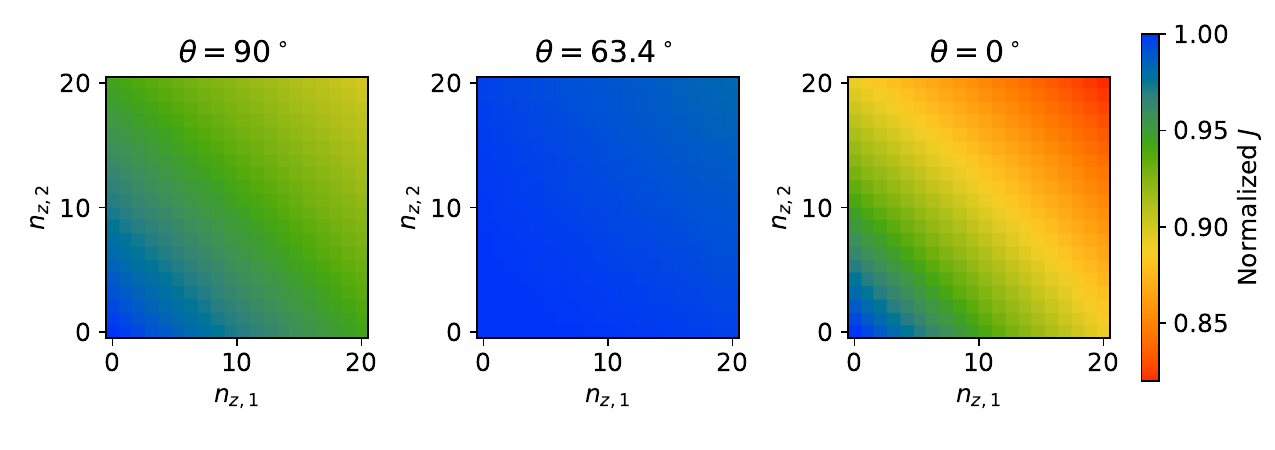}
\caption{Representative matrix slices of $J(\theta; \mathbf{n}_1; \mathbf{n}_2)/J(\theta;\mathbf{0}; \mathbf{0})$ calculated using full 3D wavefunctions. Each panel shows the interaction strength as a function of the motional quantum numbers along the $z$ axis while $\{n_{x,1}, n_{x,2}, n_{y,1}, n_{y,2}\}=0$.}
\label{fig:quantum_J_matrix}
\end{figure}

To estimate the magnitude of dephasing caused by thermal motion, we numerically calculate the matrix elements given by:
\begin{equation}
J(\mathbf{n}_1; \mathbf{n}_2)/2 = - \frac{d_{\downarrow\uparrow}^2}{4\pi\varepsilon_0} \, 
\left\langle \mathbf{n}_1 \mathbf{n}_2 \middle| 
\frac{1 - 3 \cos^2 \theta(\mathbf{r})}{|\mathbf{r}_1 - \mathbf{r}_2|^3} 
\middle| \mathbf{n}_1 \mathbf{n}_2 \right\rangle
\end{equation}
where $\ket{\mathbf{n}_i} \equiv \ket{n_{x,i}, n_{y,i}, n_{z,i}}$ denotes the motional quantum state of molecule $i$ in a 3D harmonic trap. The axes are defined as follows: $x$ is the quantization axis (external field direction), $y$ is the orthogonal radial direction, and $z$ is the direction of tweezer beam propagation (as in Fig.~1(a) of the main text).

Because $J$ is a six-dimensional tensor over motional states, we present representative slices in which the axial motional quantum state for two molecules is varied while other two directions are held fixed at $n_{x} = n_{y} = 0$. In Fig.~\ref{fig:quantum_J_matrix}, we show the $J(\theta;0,0,n_{z,1};0,0, n_{z,2})/J(\theta;0,0,0;0,0,0)$ for different $n_{z,1}, n_{z,2}$ at the $\theta = 0^\circ, 63.4^\circ$ and $90^\circ$. At the magic angle, where $A_{z,2}(63.4^\circ)=0$, the interaction strength $J$ is robust against thermal motion along $z$ direction. Simulations are performed for CaF molecules trapped in a tweezer array with axial (radial) trap frequencies of 20\,kHz (100\,kHz) and a spacing of $r=2\,\mu \mathrm{m}$.

\section{Realistic Experimental Parameters for Interaction and Rearrangement}

To assess the feasibility of dynamic geometry control in ultracold molecular tweezer arrays, we summarize key experimental parameters relevant to rearrangement speed and dipolar interaction strength for several polar molecule species. 

\vspace{0.5em}
\noindent\textbf{Dipolar Interaction Strengths.} Polar molecules possess permanent dipole moments in the molecular frame, but in the absence of an external electric field, the lab-frame static dipole moment vanishes due to rotational averaging. Nevertheless, molecules in selected rotational states can interact via dipole-mediated exchange of rotational excitations. For a given choice of interacting states, this results in a spin-exchange interaction. Table~\ref{tab:dipolar} lists representative values of the  dipole moment $d_{\mathrm{rot}}$ in the rotating molecule-frame, the effective dipole moment $d_{\mathrm{eff}}$ in the lab-frame, and the value of $J(90^\circ)$ at $r = 2~\mu\mathrm{m}$ for several species.

\begin{table}[!htbp]
\centering

\begin{tabular}{lcccc}
\toprule
\textbf{Species} & \textbf{$d_{\mathrm{rot}}$ (Debye)}& \textbf{$d_{\mathrm{lab}}$ (Debye)} & \textbf{$J$ (Hz)} & \textbf{Refs.} \\
\midrule
CaF     & $3.07$   & $1.0$   & $ 38$     & \cite{bao2023dipolar, holland2023demand} \\
NaCs    & $4.6$   & $2.7$   & $ 275$ & \cite{picard2025entanglement} \\
RbCs    & $1.2$   & $0.7$   & $ 18 $ & \cite{ruttley2025long} \\
KRb     & $0.57$   & $0.3$   & $ 3.4$     & 
\cite{yan2013observation} \\
NaRb     & $3.3$   & $1.9$   & $ 136$     & 
\cite{christakis2023probing} \\
\bottomrule
% {REFERENCES MISSING in table! and force table to go above next section.}
\end{tabular}

\caption{Realistic dipolar interaction strengths for selected polar molecules in rotational states, assuming an effective dipole moment $d_{\mathrm{lab}}$ in the lab-frame and interparticle spacing $r = 2~\mu\mathrm{m}$. The spin-exchange oscillation frequency is given by $J=2V_{dd} = d_{\mathrm{lab}}^2 / (2\pi\epsilon_0 r^3)$.}
\label{tab:dipolar}
\end{table}

\vspace{0.5em}
\noindent\textbf{Rearrangement Speed} ~~While this work focuses on molecules, insight can be drawn from atomic arrays, where high-fidelity entanglement in hyperfine ground states persists for moves slower than 200\,$\mu$s over 110\,$\mu$m, corresponding to an effective speed limit of $\sim$0.55\,$\mu$m/$\mu$s~\cite{evered2023high}. This provides a useful benchmark for estimating acceptable rearrangement conditions. Molecular tweezer arrays could operate with comparable lattice spacings and trap frequencies, allowing similar rearrangements speed. For reasonable interaction strengths of 100\,$\mathrm{Hz}$ (see Tab.~\ref{tab:dipolar}), this corresponds to $\sim 100$ moves per interaction cycle $1/J$. These timescales are well below both dipolar interaction times and coherence times, supporting dynamic geometry reshaping during entangling operations. However, while atom loss due to excessive speed has been studied, a systematic investigation of how movement contributes to thermal excitation and dephasing has yet to be performed in molecules. This question is especially relevant for molecular systems, where dipolar interactions are much weaker than in Rydberg atoms and therefore more sensitive to thermal motion.

\section{Squeezing Protocols}
While the squeezing protocol demonstrated in the main text reproduces all-to-all connectivity, there are a variety of different sets of moves that produce improved spin squeezing.  In order to faithfully reproduce an all-to-all Hamiltonian, each spin should interact equivalently with all other spins.  Starting from a spatially localized array, $O(N)$ rearrangements are necessary per cycle for each spin to interact with every other spin, producing an effective all-to-all coupling.  Ideally each cycle is repeated within a Floquet regime, where particle motion is fast in comparison to the collective dynamics.  In this section we discuss two alternative algorithms, highlighting the flexibility of rearranging particles during evolution. In practice, the following analysis holds if each move is completed on a timescale much shorter than the interaction period between moves. For more advanced protocols, one could instead employ continuous movement and account for interactions occurring during the motion.

\begin{figure}[htbp]
    \centering \includegraphics{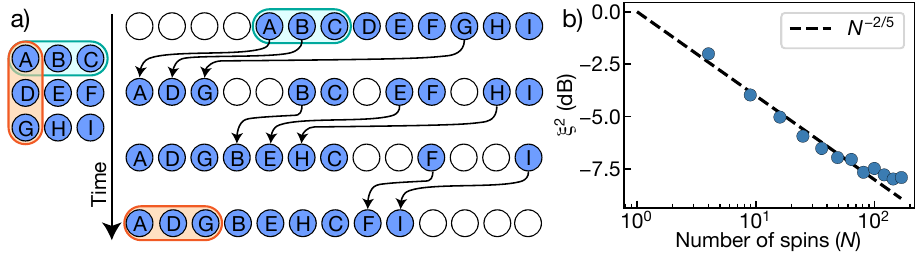}
    \caption{Rearrangement protocol for producing effective two-dimensional dynamics in a one-dimensional array.  a) A $3\times 3$ grid, highlighting a single row and column, as well as the set of rearrangements that suffice to produce effective 2D dynamics.  In this configuration the molecules interact primarily in the top and bottom configurations, while the middle two steps show a set of steps that can be performed with an acousto-optic deflector.  The corresponding row and column are highlighted.  b) Computed squeezing parameter $\xi^2$, from DTWA, for 2 cycles of the rearrangement shown in (a).  The dashed line shows $N^{-2/5},$ the predicted scaling of metrological gain for 2D dipolar arrays~\cite{block2024scalable}.}
    \label{fig:supp-sq}
\end{figure}

In the first approach we start from a one-dimensional array and reproduce an effectively $d$-dimensional array. The per-cycle number of rearrangements is then $O\left(N^{\frac{d -1}{d}}\right)$. As an illustrative example, let us consider a two-dimensional square array.  Geometrically, each molecule has 4 neighbors, two vertical and two horizontal, whereas in one dimension each molecule has just two neighbors. For $N$ molecules indexed $m_i$ in a single dimension, we perform $\sqrt{N}$ moves, each addressing $\sqrt{N}$ molecules.  On move $k$ we select all molecules $m_i$ where $i\equiv k \mod \sqrt{N}$ and rearrange such that they are adjacent.  This procedure, illustrated in Fig.~\ref{fig:supp-sq}, effectively moves molecules from interacting within a row to interacting within a column.  Since there are $\sqrt{N}$ rows that require rearranging, this procedure requires just $O(\sqrt{N})$ moves per cycle to reproduce two-dimensional squeezing.

This same approach can be used to emulate higher-dimensional arrays, at arbitrary $d$, assuming that $N^{1/d} > 1$. Each set of moves selects a slice in a different dimension, so that on a move $k$ we rearrange all molecules $m_i$ where $i\equiv k\mod N^{1/d}$ to be adjacent.  Repeating this procedure over all of the $d - 1$ additional slices results in a set of interacting pairs that mimics, up to boundary conditions, a $d$-dimensional lattice. For a $d$-dimensional array, this requires $O(N^{(d - 1)/d})$ total moves per cycle. In the limit of large $d$, we require $O(N)$ total rearrangements per cycle, and approach the limit of all-to-all couplings.

% \avikar{In a second approach, we numerically show that $O(\log N)$ rearrangements are sufficient to produce scalable spin squeezing.} 
% Taking inspiration from previous proposals for generating quantum correlations from shuffling operations, we propose a tree-like structure that produces equivalent interaction strengths at long and short distances~\cite{bentsen2019treelike, hashizume2021deterministic}.
In a second approach, we propose a tree-like structure that produces equivalent interaction strengths at long and short distances~\cite{bentsen2019treelike, hashizume2021deterministic}, with a per-cycle number of moves scaling as $O(\log N)$. Despite the relatively few moves per cycle, this protocol should still produce substantial enhancement in the achievable metrological gain, demonstrated by DTWA numerics in Fig.~\ref{fig:supp-tree}b. For $N$ molecules labeled $m_i$, we begin by performing a shuffle over the entire array, so that $m_i$ neighbors $m_{i + 2}$.  We then divide the array in half, and reapply the shuffle operation on each half of the array. We then subdivide and repeat the shuffle on each subdivision for a total of $\log_2 N - 1$ iterations.  This procedure is illustrated in Fig.~\ref{fig:supp-tree}a for an array of length 8. At each stage $i$ of this rearrangement scheme, molecule $m_j$ will interact with molecules $m_{j \pm 2^{i}}$, up to boundary effects.  These interactions at a wide range of length scales help to generate the long-distance correlations needed to produce large squeezing. 

\begin{figure}[!htbp]
    \centering
    \includegraphics{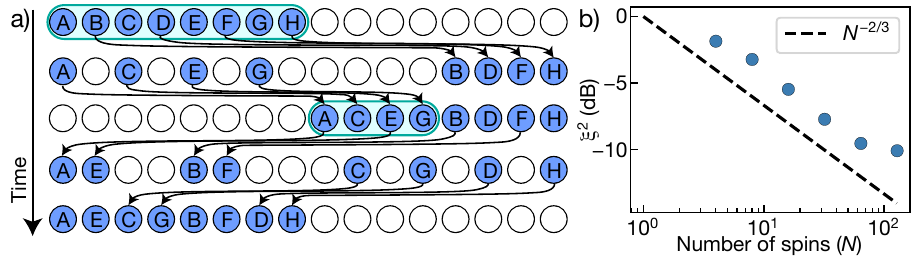}
    \caption{Logarithmic rearrangement of spins to produce enhanced spin squeezing. a) An example for 8 spins begins by shuffling the spins so that each spin interacts with its next-nearest neighbor.  We then re-shuffle the spins in each half of the array (highlighted), and repeat the process for each remaining subdivision. Interactions primarily occur in the top, middle, and bottom rows, with a full array of molecules. b) DTWA simulated squeezing parameter as a function of 1D array length, shown for 3 cycles of tree rearrangement.}
    \label{fig:supp-tree}
\end{figure}

Each shuffle operation can be performed in parallel for each subdivision of the entire array, such that the total number of required moving operations per cycle is $O(\log N)$. As with the previous example, we should note the number of cycles required to achieve maximum possible squeezing will additionally increase with $N$, and that a larger number of cycles will asymptotically approach the optimal squeezing for all of these alternative rearrangement schemes.  These schemes are designed for movement with acousto-optic deflectors.  However, recent developments in atomic motion using high-refresh rate spatial light modulators~\cite{lin2024ai} would allow arbitrary continuous rearrangement in multiple dimensions.

These alternatives are by no means a comprehensive list, and rearrangement schemes for arrays and lattices of molecules trapped in higher dimensions would also provide increased metrological gain.
% \ifSubfilesClassLoaded{%
%   \bibliographystyle{apsrev4-1}
%   \bibliography{bib}

\begin{thebibliography}{0}%
\makeatletter
\providecommand \@ifxundefined [1]{%
 \@ifx{#1\undefined}
}%
\providecommand \@ifnum [1]{%
 \ifnum #1\expandafter \@firstoftwo
 \else \expandafter \@secondoftwo
 \fi
}%
\providecommand \@ifx [1]{%
 \ifx #1\expandafter \@firstoftwo
 \else \expandafter \@secondoftwo
 \fi
}%
\providecommand \natexlab [1]{#1}%
\providecommand \enquote  [1]{``#1''}%
\providecommand \bibnamefont  [1]{#1}%
\providecommand \bibfnamefont [1]{#1}%
\providecommand \citenamefont [1]{#1}%
\providecommand \href@noop [0]{\@secondoftwo}%
\providecommand \href [0]{\begingroup \@sanitize@url \@href}%
\providecommand \@href[1]{\@@startlink{#1}\@@href}%
\providecommand \@@href[1]{\endgroup#1\@@endlink}%
\providecommand \@sanitize@url [0]{\catcode `\\12\catcode `\$12\catcode `\&12\catcode `\#12\catcode `\^12\catcode `\_12\catcode `\%12\relax}%
\providecommand \@@startlink[1]{}%
\providecommand \@@endlink[0]{}%
\providecommand \url  [0]{\begingroup\@sanitize@url \@url }%
\providecommand \@url [1]{\endgroup\@href {#1}{\urlprefix }}%
\providecommand \urlprefix  [0]{URL }%
\providecommand \Eprint [0]{\href }%
\providecommand \doibase [0]{https://doi.org/}%
\providecommand \selectlanguage [0]{\@gobble}%
\providecommand \bibinfo  [0]{\@secondoftwo}%
\providecommand \bibfield  [0]{\@secondoftwo}%
\providecommand \translation [1]{[#1]}%
\providecommand \BibitemOpen [0]{}%
\providecommand \bibitemStop [0]{}%
\providecommand \bibitemNoStop [0]{.\EOS\space}%
\providecommand \EOS [0]{\spacefactor3000\relax}%
\providecommand \BibitemShut  [1]{\csname bibitem#1\endcsname}%
\let\auto@bib@innerbib\@empty
%</preamble>
\end{thebibliography}%


\providecommand{\noopsort}[1]{}\providecommand{\singleletter}[1]{#1}%
\begin{thebibliography}{44}%
\makeatletter
\providecommand \@ifxundefined [1]{%
 \@ifx{#1\undefined}
}%
\providecommand \@ifnum [1]{%
 \ifnum #1\expandafter \@firstoftwo
 \else \expandafter \@secondoftwo
 \fi
}%
\providecommand \@ifx [1]{%
 \ifx #1\expandafter \@firstoftwo
 \else \expandafter \@secondoftwo
 \fi
}%
\providecommand \natexlab [1]{#1}%
\providecommand \enquote  [1]{``#1''}%
\providecommand \bibnamefont  [1]{#1}%
\providecommand \bibfnamefont [1]{#1}%
\providecommand \citenamefont [1]{#1}%
\providecommand \href@noop [0]{\@secondoftwo}%
\providecommand \href [0]{\begingroup \@sanitize@url \@href}%
\providecommand \@href[1]{\@@startlink{#1}\@@href}%
\providecommand \@@href[1]{\endgroup#1\@@endlink}%
\providecommand \@sanitize@url [0]{\catcode `\\12\catcode `\$12\catcode `\&12\catcode `\#12\catcode `\^12\catcode `\_12\catcode `\%12\relax}%
\providecommand \@@startlink[1]{}%
\providecommand \@@endlink[0]{}%
\providecommand \url  [0]{\begingroup\@sanitize@url \@url }%
\providecommand \@url [1]{\endgroup\@href {#1}{\urlprefix }}%
\providecommand \urlprefix  [0]{URL }%
\providecommand \Eprint [0]{\href }%
\providecommand \doibase [0]{http://dx.doi.org/}%
\providecommand \selectlanguage [0]{\@gobble}%
\providecommand \bibinfo  [0]{\@secondoftwo}%
\providecommand \bibfield  [0]{\@secondoftwo}%
\providecommand \translation [1]{[#1]}%
\providecommand \BibitemOpen [0]{}%
\providecommand \bibitemStop [0]{}%
\providecommand \bibitemNoStop [0]{.\EOS\space}%
\providecommand \EOS [0]{\spacefactor3000\relax}%
\providecommand \BibitemShut  [1]{\csname bibitem#1\endcsname}%
\let\auto@bib@innerbib\@empty
%</preamble>
\bibitem [{\citenamefont {Burchesky}\ \emph {et~al.}(2021)\citenamefont {Burchesky}, \citenamefont {Anderegg}, \citenamefont {Bao}, \citenamefont {Yu}, \citenamefont {Chae}, \citenamefont {Ketterle}, \citenamefont {Ni},\ and\ \citenamefont {Doyle}}]{burchesky2021rotational}%
  \BibitemOpen
  \bibfield  {author} {\bibinfo {author} {\bibfnamefont {S.}~\bibnamefont {Burchesky}}, \bibinfo {author} {\bibfnamefont {L.}~\bibnamefont {Anderegg}}, \bibinfo {author} {\bibfnamefont {Y.}~\bibnamefont {Bao}}, \bibinfo {author} {\bibfnamefont {S.~S.}\ \bibnamefont {Yu}}, \bibinfo {author} {\bibfnamefont {E.}~\bibnamefont {Chae}}, \bibinfo {author} {\bibfnamefont {W.}~\bibnamefont {Ketterle}}, \bibinfo {author} {\bibfnamefont {K.-K.}\ \bibnamefont {Ni}}, \ and\ \bibinfo {author} {\bibfnamefont {J.~M.}\ \bibnamefont {Doyle}},\ }\href@noop {} {\bibfield  {journal} {\bibinfo  {journal} {Physical Review Letters}\ }\textbf {\bibinfo {volume} {127}},\ \bibinfo {pages} {123202} (\bibinfo {year} {2021})}\BibitemShut {NoStop}%
\bibitem [{\citenamefont {Park}\ \emph {et~al.}(2023)\citenamefont {Park}, \citenamefont {Picard}, \citenamefont {Patenotte}, \citenamefont {Zhang}, \citenamefont {Rosenband},\ and\ \citenamefont {Ni}}]{park2023extended}%
  \BibitemOpen
  \bibfield  {author} {\bibinfo {author} {\bibfnamefont {A.~J.}\ \bibnamefont {Park}}, \bibinfo {author} {\bibfnamefont {L.~R.}\ \bibnamefont {Picard}}, \bibinfo {author} {\bibfnamefont {G.~E.}\ \bibnamefont {Patenotte}}, \bibinfo {author} {\bibfnamefont {J.~T.}\ \bibnamefont {Zhang}}, \bibinfo {author} {\bibfnamefont {T.}~\bibnamefont {Rosenband}}, \ and\ \bibinfo {author} {\bibfnamefont {K.-K.}\ \bibnamefont {Ni}},\ }\href@noop {} {\bibfield  {journal} {\bibinfo  {journal} {Physical Review Letters}\ }\textbf {\bibinfo {volume} {131}},\ \bibinfo {pages} {183401} (\bibinfo {year} {2023})}\BibitemShut {NoStop}%
\bibitem [{\citenamefont {Gregory}\ \emph {et~al.}(2024)\citenamefont {Gregory}, \citenamefont {Fernley}, \citenamefont {Tao}, \citenamefont {Bromley}, \citenamefont {Stepp}, \citenamefont {Zhang}, \citenamefont {Kotochigova}, \citenamefont {Hazzard},\ and\ \citenamefont {Cornish}}]{gregory2024second}%
  \BibitemOpen
  \bibfield  {author} {\bibinfo {author} {\bibfnamefont {P.~D.}\ \bibnamefont {Gregory}}, \bibinfo {author} {\bibfnamefont {L.~M.}\ \bibnamefont {Fernley}}, \bibinfo {author} {\bibfnamefont {A.~L.}\ \bibnamefont {Tao}}, \bibinfo {author} {\bibfnamefont {S.~L.}\ \bibnamefont {Bromley}}, \bibinfo {author} {\bibfnamefont {J.}~\bibnamefont {Stepp}}, \bibinfo {author} {\bibfnamefont {Z.}~\bibnamefont {Zhang}}, \bibinfo {author} {\bibfnamefont {S.}~\bibnamefont {Kotochigova}}, \bibinfo {author} {\bibfnamefont {K.~R.}\ \bibnamefont {Hazzard}}, \ and\ \bibinfo {author} {\bibfnamefont {S.~L.}\ \bibnamefont {Cornish}},\ }\href@noop {} {\bibfield  {journal} {\bibinfo  {journal} {Nature Physics}\ }\textbf {\bibinfo {volume} {20}},\ \bibinfo {pages} {415} (\bibinfo {year} {2024})}\BibitemShut {NoStop}%
\bibitem [{\citenamefont {Ni}\ \emph {et~al.}(2018)\citenamefont {Ni}, \citenamefont {Rosenband},\ and\ \citenamefont {Grimes}}]{ni2018dipolar}%
  \BibitemOpen
  \bibfield  {author} {\bibinfo {author} {\bibfnamefont {K.-K.}\ \bibnamefont {Ni}}, \bibinfo {author} {\bibfnamefont {T.}~\bibnamefont {Rosenband}}, \ and\ \bibinfo {author} {\bibfnamefont {D.~D.}\ \bibnamefont {Grimes}},\ }\href@noop {} {\bibfield  {journal} {\bibinfo  {journal} {Chemical science}\ }\textbf {\bibinfo {volume} {9}},\ \bibinfo {pages} {6830} (\bibinfo {year} {2018})}\BibitemShut {NoStop}%
\bibitem [{\citenamefont {Holland}\ \emph {et~al.}(2023)\citenamefont {Holland}, \citenamefont {Lu},\ and\ \citenamefont {Cheuk}}]{holland2023demand}%
  \BibitemOpen
  \bibfield  {author} {\bibinfo {author} {\bibfnamefont {C.~M.}\ \bibnamefont {Holland}}, \bibinfo {author} {\bibfnamefont {Y.}~\bibnamefont {Lu}}, \ and\ \bibinfo {author} {\bibfnamefont {L.~W.}\ \bibnamefont {Cheuk}},\ }\href@noop {} {\bibfield  {journal} {\bibinfo  {journal} {Science}\ }\textbf {\bibinfo {volume} {382}},\ \bibinfo {pages} {1143} (\bibinfo {year} {2023})}\BibitemShut {NoStop}%
\bibitem [{\citenamefont {Picard}\ \emph {et~al.}(2025)\citenamefont {Picard}, \citenamefont {Park}, \citenamefont {Patenotte}, \citenamefont {Gebretsadkan}, \citenamefont {Wellnitz}, \citenamefont {Rey},\ and\ \citenamefont {Ni}}]{picard2025entanglement}%
  \BibitemOpen
  \bibfield  {author} {\bibinfo {author} {\bibfnamefont {L.~R.}\ \bibnamefont {Picard}}, \bibinfo {author} {\bibfnamefont {A.~J.}\ \bibnamefont {Park}}, \bibinfo {author} {\bibfnamefont {G.~E.}\ \bibnamefont {Patenotte}}, \bibinfo {author} {\bibfnamefont {S.}~\bibnamefont {Gebretsadkan}}, \bibinfo {author} {\bibfnamefont {D.}~\bibnamefont {Wellnitz}}, \bibinfo {author} {\bibfnamefont {A.~M.}\ \bibnamefont {Rey}}, \ and\ \bibinfo {author} {\bibfnamefont {K.-K.}\ \bibnamefont {Ni}},\ }\href@noop {} {\bibfield  {journal} {\bibinfo  {journal} {Nature}\ }\textbf {\bibinfo {volume} {637}},\ \bibinfo {pages} {821} (\bibinfo {year} {2025})}\BibitemShut {NoStop}%
\bibitem [{\citenamefont {Hepworth}\ \emph {et~al.}(2024)\citenamefont {Hepworth}, \citenamefont {Ruttley}, \citenamefont {von Gierke}, \citenamefont {Gregory}, \citenamefont {Guttridge},\ and\ \citenamefont {Cornish}}]{hepworth2024coherent}%
  \BibitemOpen
  \bibfield  {author} {\bibinfo {author} {\bibfnamefont {T.~R.}\ \bibnamefont {Hepworth}}, \bibinfo {author} {\bibfnamefont {D.~K.}\ \bibnamefont {Ruttley}}, \bibinfo {author} {\bibfnamefont {F.}~\bibnamefont {von Gierke}}, \bibinfo {author} {\bibfnamefont {P.~D.}\ \bibnamefont {Gregory}}, \bibinfo {author} {\bibfnamefont {A.}~\bibnamefont {Guttridge}}, \ and\ \bibinfo {author} {\bibfnamefont {S.~L.}\ \bibnamefont {Cornish}},\ }\href@noop {} {\bibfield  {journal} {\bibinfo  {journal} {arXiv preprint arXiv:2412.15088}\ } (\bibinfo {year} {2024})}\BibitemShut {NoStop}%
\bibitem [{\citenamefont {Anderegg}\ \emph {et~al.}(2019)\citenamefont {Anderegg}, \citenamefont {Cheuk}, \citenamefont {Bao}, \citenamefont {Burchesky}, \citenamefont {Ketterle}, \citenamefont {Ni},\ and\ \citenamefont {Doyle}}]{anderegg2019optical}%
  \BibitemOpen
  \bibfield  {author} {\bibinfo {author} {\bibfnamefont {L.}~\bibnamefont {Anderegg}}, \bibinfo {author} {\bibfnamefont {L.~W.}\ \bibnamefont {Cheuk}}, \bibinfo {author} {\bibfnamefont {Y.}~\bibnamefont {Bao}}, \bibinfo {author} {\bibfnamefont {S.}~\bibnamefont {Burchesky}}, \bibinfo {author} {\bibfnamefont {W.}~\bibnamefont {Ketterle}}, \bibinfo {author} {\bibfnamefont {K.-K.}\ \bibnamefont {Ni}}, \ and\ \bibinfo {author} {\bibfnamefont {J.~M.}\ \bibnamefont {Doyle}},\ }\href@noop {} {\bibfield  {journal} {\bibinfo  {journal} {Science}\ }\textbf {\bibinfo {volume} {365}},\ \bibinfo {pages} {1156} (\bibinfo {year} {2019})}\BibitemShut {NoStop}%
\bibitem [{\citenamefont {Vilas}\ \emph {et~al.}(2024)\citenamefont {Vilas}, \citenamefont {Robichaud}, \citenamefont {Hallas}, \citenamefont {Li}, \citenamefont {Anderegg},\ and\ \citenamefont {Doyle}}]{vilas2024optical}%
  \BibitemOpen
  \bibfield  {author} {\bibinfo {author} {\bibfnamefont {N.~B.}\ \bibnamefont {Vilas}}, \bibinfo {author} {\bibfnamefont {P.}~\bibnamefont {Robichaud}}, \bibinfo {author} {\bibfnamefont {C.}~\bibnamefont {Hallas}}, \bibinfo {author} {\bibfnamefont {G.~K.}\ \bibnamefont {Li}}, \bibinfo {author} {\bibfnamefont {L.}~\bibnamefont {Anderegg}}, \ and\ \bibinfo {author} {\bibfnamefont {J.~M.}\ \bibnamefont {Doyle}},\ }\href@noop {} {\bibfield  {journal} {\bibinfo  {journal} {Nature}\ }\textbf {\bibinfo {volume} {628}},\ \bibinfo {pages} {282} (\bibinfo {year} {2024})}\BibitemShut {NoStop}%
\bibitem [{\citenamefont {Park}\ \emph {et~al.}(2017)\citenamefont {Park}, \citenamefont {Yan}, \citenamefont {Loh}, \citenamefont {Will},\ and\ \citenamefont {Zwierlein}}]{park2017second}%
  \BibitemOpen
  \bibfield  {author} {\bibinfo {author} {\bibfnamefont {J.~W.}\ \bibnamefont {Park}}, \bibinfo {author} {\bibfnamefont {Z.~Z.}\ \bibnamefont {Yan}}, \bibinfo {author} {\bibfnamefont {H.}~\bibnamefont {Loh}}, \bibinfo {author} {\bibfnamefont {S.~A.}\ \bibnamefont {Will}}, \ and\ \bibinfo {author} {\bibfnamefont {M.~W.}\ \bibnamefont {Zwierlein}},\ }\href@noop {} {\bibfield  {journal} {\bibinfo  {journal} {Science}\ }\textbf {\bibinfo {volume} {357}},\ \bibinfo {pages} {372} (\bibinfo {year} {2017})}\BibitemShut {NoStop}%
\bibitem [{\citenamefont {Ruttley}\ \emph {et~al.}(2025)\citenamefont {Ruttley}, \citenamefont {Hepworth}, \citenamefont {Guttridge},\ and\ \citenamefont {Cornish}}]{ruttley2025long}%
  \BibitemOpen
  \bibfield  {author} {\bibinfo {author} {\bibfnamefont {D.~K.}\ \bibnamefont {Ruttley}}, \bibinfo {author} {\bibfnamefont {T.~R.}\ \bibnamefont {Hepworth}}, \bibinfo {author} {\bibfnamefont {A.}~\bibnamefont {Guttridge}}, \ and\ \bibinfo {author} {\bibfnamefont {S.~L.}\ \bibnamefont {Cornish}},\ }\href@noop {} {\bibfield  {journal} {\bibinfo  {journal} {Nature}\ ,\ \bibinfo {pages} {1}} (\bibinfo {year} {2025})}\BibitemShut {NoStop}%
\bibitem [{\citenamefont {Silveri}\ \emph {et~al.}(2017)\citenamefont {Silveri}, \citenamefont {Tuorila}, \citenamefont {Thuneberg},\ and\ \citenamefont {Paraoanu}}]{silveri2017quantum}%
  \BibitemOpen
  \bibfield  {author} {\bibinfo {author} {\bibfnamefont {M.}~\bibnamefont {Silveri}}, \bibinfo {author} {\bibfnamefont {J.}~\bibnamefont {Tuorila}}, \bibinfo {author} {\bibfnamefont {E.}~\bibnamefont {Thuneberg}}, \ and\ \bibinfo {author} {\bibfnamefont {G.}~\bibnamefont {Paraoanu}},\ }\href@noop {} {\bibfield  {journal} {\bibinfo  {journal} {Reports on Progress in Physics}\ }\textbf {\bibinfo {volume} {80}},\ \bibinfo {pages} {056002} (\bibinfo {year} {2017})}\BibitemShut {NoStop}%
\bibitem [{\citenamefont {Blackmore}\ \emph {et~al.}(2020)\citenamefont {Blackmore}, \citenamefont {Sawant}, \citenamefont {Gregory}, \citenamefont {Bromley}, \citenamefont {Aldegunde}, \citenamefont {Hutson},\ and\ \citenamefont {Cornish}}]{blackmore2020controlling}%
  \BibitemOpen
  \bibfield  {author} {\bibinfo {author} {\bibfnamefont {J.~A.}\ \bibnamefont {Blackmore}}, \bibinfo {author} {\bibfnamefont {R.}~\bibnamefont {Sawant}}, \bibinfo {author} {\bibfnamefont {P.~D.}\ \bibnamefont {Gregory}}, \bibinfo {author} {\bibfnamefont {S.~L.}\ \bibnamefont {Bromley}}, \bibinfo {author} {\bibfnamefont {J.}~\bibnamefont {Aldegunde}}, \bibinfo {author} {\bibfnamefont {J.~M.}\ \bibnamefont {Hutson}}, \ and\ \bibinfo {author} {\bibfnamefont {S.~L.}\ \bibnamefont {Cornish}},\ }\href@noop {} {\bibfield  {journal} {\bibinfo  {journal} {Physical Review A}\ }\textbf {\bibinfo {volume} {102}},\ \bibinfo {pages} {053316} (\bibinfo {year} {2020})}\BibitemShut {NoStop}%
\bibitem [{\citenamefont {Barredo}\ \emph {et~al.}(2016)\citenamefont {Barredo}, \citenamefont {De~L{\'e}s{\'e}leuc}, \citenamefont {Lienhard}, \citenamefont {Lahaye},\ and\ \citenamefont {Browaeys}}]{barredo2016atom}%
  \BibitemOpen
  \bibfield  {author} {\bibinfo {author} {\bibfnamefont {D.}~\bibnamefont {Barredo}}, \bibinfo {author} {\bibfnamefont {S.}~\bibnamefont {De~L{\'e}s{\'e}leuc}}, \bibinfo {author} {\bibfnamefont {V.}~\bibnamefont {Lienhard}}, \bibinfo {author} {\bibfnamefont {T.}~\bibnamefont {Lahaye}}, \ and\ \bibinfo {author} {\bibfnamefont {A.}~\bibnamefont {Browaeys}},\ }\href@noop {} {\bibfield  {journal} {\bibinfo  {journal} {Science}\ }\textbf {\bibinfo {volume} {354}},\ \bibinfo {pages} {1021} (\bibinfo {year} {2016})}\BibitemShut {NoStop}%
\bibitem [{\citenamefont {Endres}\ \emph {et~al.}(2016)\citenamefont {Endres}, \citenamefont {Bernien}, \citenamefont {Keesling}, \citenamefont {Levine}, \citenamefont {Anschuetz}, \citenamefont {Krajenbrink}, \citenamefont {Senko}, \citenamefont {Vuletic}, \citenamefont {Greiner},\ and\ \citenamefont {Lukin}}]{endres2016atom}%
  \BibitemOpen
  \bibfield  {author} {\bibinfo {author} {\bibfnamefont {M.}~\bibnamefont {Endres}}, \bibinfo {author} {\bibfnamefont {H.}~\bibnamefont {Bernien}}, \bibinfo {author} {\bibfnamefont {A.}~\bibnamefont {Keesling}}, \bibinfo {author} {\bibfnamefont {H.}~\bibnamefont {Levine}}, \bibinfo {author} {\bibfnamefont {E.~R.}\ \bibnamefont {Anschuetz}}, \bibinfo {author} {\bibfnamefont {A.}~\bibnamefont {Krajenbrink}}, \bibinfo {author} {\bibfnamefont {C.}~\bibnamefont {Senko}}, \bibinfo {author} {\bibfnamefont {V.}~\bibnamefont {Vuletic}}, \bibinfo {author} {\bibfnamefont {M.}~\bibnamefont {Greiner}}, \ and\ \bibinfo {author} {\bibfnamefont {M.~D.}\ \bibnamefont {Lukin}},\ }\href@noop {} {\bibfield  {journal} {\bibinfo  {journal} {Science}\ }\textbf {\bibinfo {volume} {354}},\ \bibinfo {pages} {1024} (\bibinfo {year} {2016})}\BibitemShut {NoStop}%
\bibitem [{\citenamefont {Bluvstein}\ \emph {et~al.}(2022)\citenamefont {Bluvstein}, \citenamefont {Levine}, \citenamefont {Semeghini}, \citenamefont {Wang}, \citenamefont {Ebadi}, \citenamefont {Kalinowski}, \citenamefont {Keesling}, \citenamefont {Maskara}, \citenamefont {Pichler}, \citenamefont {Greiner} \emph {et~al.}}]{bluvstein2022quantum}%
  \BibitemOpen
  \bibfield  {author} {\bibinfo {author} {\bibfnamefont {D.}~\bibnamefont {Bluvstein}}, \bibinfo {author} {\bibfnamefont {H.}~\bibnamefont {Levine}}, \bibinfo {author} {\bibfnamefont {G.}~\bibnamefont {Semeghini}}, \bibinfo {author} {\bibfnamefont {T.~T.}\ \bibnamefont {Wang}}, \bibinfo {author} {\bibfnamefont {S.}~\bibnamefont {Ebadi}}, \bibinfo {author} {\bibfnamefont {M.}~\bibnamefont {Kalinowski}}, \bibinfo {author} {\bibfnamefont {A.}~\bibnamefont {Keesling}}, \bibinfo {author} {\bibfnamefont {N.}~\bibnamefont {Maskara}}, \bibinfo {author} {\bibfnamefont {H.}~\bibnamefont {Pichler}}, \bibinfo {author} {\bibfnamefont {M.}~\bibnamefont {Greiner}},  \emph {et~al.},\ }\href@noop {} {\bibfield  {journal} {\bibinfo  {journal} {Nature}\ }\textbf {\bibinfo {volume} {604}},\ \bibinfo {pages} {451} (\bibinfo {year} {2022})}\BibitemShut {NoStop}%
\bibitem [{\citenamefont {Deist}\ \emph {et~al.}(2022)\citenamefont {Deist}, \citenamefont {Gerber}, \citenamefont {Lu}, \citenamefont {Zeiher},\ and\ \citenamefont {Stamper-Kurn}}]{deist2022superresolution}%
  \BibitemOpen
  \bibfield  {author} {\bibinfo {author} {\bibfnamefont {E.}~\bibnamefont {Deist}}, \bibinfo {author} {\bibfnamefont {J.~A.}\ \bibnamefont {Gerber}}, \bibinfo {author} {\bibfnamefont {Y.-H.}\ \bibnamefont {Lu}}, \bibinfo {author} {\bibfnamefont {J.}~\bibnamefont {Zeiher}}, \ and\ \bibinfo {author} {\bibfnamefont {D.~M.}\ \bibnamefont {Stamper-Kurn}},\ }\href@noop {} {\bibfield  {journal} {\bibinfo  {journal} {Physical Review Letters}\ }\textbf {\bibinfo {volume} {128}},\ \bibinfo {pages} {083201} (\bibinfo {year} {2022})}\BibitemShut {NoStop}%
\bibitem [{\citenamefont {Davis}\ \emph {et~al.}(2023)\citenamefont {Davis}, \citenamefont {Ye}, \citenamefont {Machado}, \citenamefont {Meynell}, \citenamefont {Wu}, \citenamefont {Mittiga}, \citenamefont {Schenken}, \citenamefont {Joos}, \citenamefont {Kobrin}, \citenamefont {Lyu} \emph {et~al.}}]{davis2023probing}%
  \BibitemOpen
  \bibfield  {author} {\bibinfo {author} {\bibfnamefont {E.~J.}\ \bibnamefont {Davis}}, \bibinfo {author} {\bibfnamefont {B.}~\bibnamefont {Ye}}, \bibinfo {author} {\bibfnamefont {F.}~\bibnamefont {Machado}}, \bibinfo {author} {\bibfnamefont {S.~A.}\ \bibnamefont {Meynell}}, \bibinfo {author} {\bibfnamefont {W.}~\bibnamefont {Wu}}, \bibinfo {author} {\bibfnamefont {T.}~\bibnamefont {Mittiga}}, \bibinfo {author} {\bibfnamefont {W.}~\bibnamefont {Schenken}}, \bibinfo {author} {\bibfnamefont {M.}~\bibnamefont {Joos}}, \bibinfo {author} {\bibfnamefont {B.}~\bibnamefont {Kobrin}}, \bibinfo {author} {\bibfnamefont {Y.}~\bibnamefont {Lyu}},  \emph {et~al.},\ }\href@noop {} {\bibfield  {journal} {\bibinfo  {journal} {Nature physics}\ }\textbf {\bibinfo {volume} {19}},\ \bibinfo {pages} {836} (\bibinfo {year} {2023})}\BibitemShut {NoStop}%
\bibitem [{\citenamefont {Browaeys}\ and\ \citenamefont {Lahaye}(2020)}]{browaeys2020many}%
  \BibitemOpen
  \bibfield  {author} {\bibinfo {author} {\bibfnamefont {A.}~\bibnamefont {Browaeys}}\ and\ \bibinfo {author} {\bibfnamefont {T.}~\bibnamefont {Lahaye}},\ }\href@noop {} {\bibfield  {journal} {\bibinfo  {journal} {Nature Physics}\ }\textbf {\bibinfo {volume} {16}},\ \bibinfo {pages} {132} (\bibinfo {year} {2020})}\BibitemShut {NoStop}%
\bibitem [{\citenamefont {Bornet}\ \emph {et~al.}(2023)\citenamefont {Bornet}, \citenamefont {Emperauger}, \citenamefont {Chen}, \citenamefont {Ye}, \citenamefont {Block}, \citenamefont {Bintz}, \citenamefont {Boyd}, \citenamefont {Barredo}, \citenamefont {Comparin}, \citenamefont {Mezzacapo} \emph {et~al.}}]{bornet2023scalable}%
  \BibitemOpen
  \bibfield  {author} {\bibinfo {author} {\bibfnamefont {G.}~\bibnamefont {Bornet}}, \bibinfo {author} {\bibfnamefont {G.}~\bibnamefont {Emperauger}}, \bibinfo {author} {\bibfnamefont {C.}~\bibnamefont {Chen}}, \bibinfo {author} {\bibfnamefont {B.}~\bibnamefont {Ye}}, \bibinfo {author} {\bibfnamefont {M.}~\bibnamefont {Block}}, \bibinfo {author} {\bibfnamefont {M.}~\bibnamefont {Bintz}}, \bibinfo {author} {\bibfnamefont {J.~A.}\ \bibnamefont {Boyd}}, \bibinfo {author} {\bibfnamefont {D.}~\bibnamefont {Barredo}}, \bibinfo {author} {\bibfnamefont {T.}~\bibnamefont {Comparin}}, \bibinfo {author} {\bibfnamefont {F.}~\bibnamefont {Mezzacapo}},  \emph {et~al.},\ }\href@noop {} {\bibfield  {journal} {\bibinfo  {journal} {Nature}\ }\textbf {\bibinfo {volume} {621}},\ \bibinfo {pages} {728} (\bibinfo {year} {2023})}\BibitemShut {NoStop}%
\bibitem [{\citenamefont {Su}\ \emph {et~al.}(2023)\citenamefont {Su}, \citenamefont {Douglas}, \citenamefont {Szurek}, \citenamefont {Groth}, \citenamefont {Ozturk}, \citenamefont {Krahn}, \citenamefont {H{\'e}bert}, \citenamefont {Phelps}, \citenamefont {Ebadi}, \citenamefont {Dickerson} \emph {et~al.}}]{su2023dipolar}%
  \BibitemOpen
  \bibfield  {author} {\bibinfo {author} {\bibfnamefont {L.}~\bibnamefont {Su}}, \bibinfo {author} {\bibfnamefont {A.}~\bibnamefont {Douglas}}, \bibinfo {author} {\bibfnamefont {M.}~\bibnamefont {Szurek}}, \bibinfo {author} {\bibfnamefont {R.}~\bibnamefont {Groth}}, \bibinfo {author} {\bibfnamefont {S.~F.}\ \bibnamefont {Ozturk}}, \bibinfo {author} {\bibfnamefont {A.}~\bibnamefont {Krahn}}, \bibinfo {author} {\bibfnamefont {A.~H.}\ \bibnamefont {H{\'e}bert}}, \bibinfo {author} {\bibfnamefont {G.~A.}\ \bibnamefont {Phelps}}, \bibinfo {author} {\bibfnamefont {S.}~\bibnamefont {Ebadi}}, \bibinfo {author} {\bibfnamefont {S.}~\bibnamefont {Dickerson}},  \emph {et~al.},\ }\href@noop {} {\bibfield  {journal} {\bibinfo  {journal} {Nature}\ }\textbf {\bibinfo {volume} {622}},\ \bibinfo {pages} {724} (\bibinfo {year} {2023})}\BibitemShut {NoStop}%
\bibitem [{\citenamefont {Raeisi}\ \emph {et~al.}(2012)\citenamefont {Raeisi}, \citenamefont {Wiebe},\ and\ \citenamefont {Sanders}}]{raeisi2012quantum}%
  \BibitemOpen
  \bibfield  {author} {\bibinfo {author} {\bibfnamefont {S.}~\bibnamefont {Raeisi}}, \bibinfo {author} {\bibfnamefont {N.}~\bibnamefont {Wiebe}}, \ and\ \bibinfo {author} {\bibfnamefont {B.~C.}\ \bibnamefont {Sanders}},\ }\href@noop {} {\bibfield  {journal} {\bibinfo  {journal} {New Journal of Physics}\ }\textbf {\bibinfo {volume} {14}},\ \bibinfo {pages} {103017} (\bibinfo {year} {2012})}\BibitemShut {NoStop}%
\bibitem [{\citenamefont {Bluvstein}\ \emph {et~al.}(2024)\citenamefont {Bluvstein}, \citenamefont {Evered}, \citenamefont {Geim}, \citenamefont {Li}, \citenamefont {Zhou}, \citenamefont {Manovitz}, \citenamefont {Ebadi}, \citenamefont {Cain}, \citenamefont {Kalinowski}, \citenamefont {Hangleiter} \emph {et~al.}}]{bluvstein2024logical}%
  \BibitemOpen
  \bibfield  {author} {\bibinfo {author} {\bibfnamefont {D.}~\bibnamefont {Bluvstein}}, \bibinfo {author} {\bibfnamefont {S.~J.}\ \bibnamefont {Evered}}, \bibinfo {author} {\bibfnamefont {A.~A.}\ \bibnamefont {Geim}}, \bibinfo {author} {\bibfnamefont {S.~H.}\ \bibnamefont {Li}}, \bibinfo {author} {\bibfnamefont {H.}~\bibnamefont {Zhou}}, \bibinfo {author} {\bibfnamefont {T.}~\bibnamefont {Manovitz}}, \bibinfo {author} {\bibfnamefont {S.}~\bibnamefont {Ebadi}}, \bibinfo {author} {\bibfnamefont {M.}~\bibnamefont {Cain}}, \bibinfo {author} {\bibfnamefont {M.}~\bibnamefont {Kalinowski}}, \bibinfo {author} {\bibfnamefont {D.}~\bibnamefont {Hangleiter}},  \emph {et~al.},\ }\href@noop {} {\bibfield  {journal} {\bibinfo  {journal} {Nature}\ }\textbf {\bibinfo {volume} {626}},\ \bibinfo {pages} {58} (\bibinfo {year} {2024})}\BibitemShut {NoStop}%
\bibitem [{\citenamefont {Bao}\ \emph {et~al.}(2023)\citenamefont {Bao}, \citenamefont {Yu}, \citenamefont {Anderegg}, \citenamefont {Chae}, \citenamefont {Ketterle}, \citenamefont {Ni},\ and\ \citenamefont {Doyle}}]{bao2023dipolar}%
  \BibitemOpen
  \bibfield  {author} {\bibinfo {author} {\bibfnamefont {Y.}~\bibnamefont {Bao}}, \bibinfo {author} {\bibfnamefont {S.~S.}\ \bibnamefont {Yu}}, \bibinfo {author} {\bibfnamefont {L.}~\bibnamefont {Anderegg}}, \bibinfo {author} {\bibfnamefont {E.}~\bibnamefont {Chae}}, \bibinfo {author} {\bibfnamefont {W.}~\bibnamefont {Ketterle}}, \bibinfo {author} {\bibfnamefont {K.-K.}\ \bibnamefont {Ni}}, \ and\ \bibinfo {author} {\bibfnamefont {J.~M.}\ \bibnamefont {Doyle}},\ }\href@noop {} {\bibfield  {journal} {\bibinfo  {journal} {Science}\ }\textbf {\bibinfo {volume} {382}},\ \bibinfo {pages} {1138} (\bibinfo {year} {2023})}\BibitemShut {NoStop}%
\bibitem [{\citenamefont {Bergonzoni}\ \emph {et~al.}(2025)\citenamefont {Bergonzoni}, \citenamefont {Jandura},\ and\ \citenamefont {Pupillo}}]{bergonzoni2025iswap}%
  \BibitemOpen
  \bibfield  {author} {\bibinfo {author} {\bibfnamefont {M.}~\bibnamefont {Bergonzoni}}, \bibinfo {author} {\bibfnamefont {S.}~\bibnamefont {Jandura}}, \ and\ \bibinfo {author} {\bibfnamefont {G.}~\bibnamefont {Pupillo}},\ }\href@noop {} {\bibfield  {journal} {\bibinfo  {journal} {arXiv preprint arXiv:2502.21238}\ } (\bibinfo {year} {2025})}\BibitemShut {NoStop}%
\bibitem [{\citenamefont {Anderegg}\ \emph {et~al.}(2018)\citenamefont {Anderegg}, \citenamefont {Augenbraun}, \citenamefont {Bao}, \citenamefont {Burchesky}, \citenamefont {Cheuk}, \citenamefont {Ketterle},\ and\ \citenamefont {Doyle}}]{anderegg2018laser}%
  \BibitemOpen
  \bibfield  {author} {\bibinfo {author} {\bibfnamefont {L.}~\bibnamefont {Anderegg}}, \bibinfo {author} {\bibfnamefont {B.~L.}\ \bibnamefont {Augenbraun}}, \bibinfo {author} {\bibfnamefont {Y.}~\bibnamefont {Bao}}, \bibinfo {author} {\bibfnamefont {S.}~\bibnamefont {Burchesky}}, \bibinfo {author} {\bibfnamefont {L.~W.}\ \bibnamefont {Cheuk}}, \bibinfo {author} {\bibfnamefont {W.}~\bibnamefont {Ketterle}}, \ and\ \bibinfo {author} {\bibfnamefont {J.~M.}\ \bibnamefont {Doyle}},\ }\href@noop {} {\bibfield  {journal} {\bibinfo  {journal} {Nature Physics}\ }\textbf {\bibinfo {volume} {14}},\ \bibinfo {pages} {890} (\bibinfo {year} {2018})}\BibitemShut {NoStop}%
\bibitem [{\citenamefont {Cheuk}\ \emph {et~al.}(2018)\citenamefont {Cheuk}, \citenamefont {Anderegg}, \citenamefont {Augenbraun}, \citenamefont {Bao}, \citenamefont {Burchesky}, \citenamefont {Ketterle},\ and\ \citenamefont {Doyle}}]{cheuk2018lambda}%
  \BibitemOpen
  \bibfield  {author} {\bibinfo {author} {\bibfnamefont {L.~W.}\ \bibnamefont {Cheuk}}, \bibinfo {author} {\bibfnamefont {L.}~\bibnamefont {Anderegg}}, \bibinfo {author} {\bibfnamefont {B.~L.}\ \bibnamefont {Augenbraun}}, \bibinfo {author} {\bibfnamefont {Y.}~\bibnamefont {Bao}}, \bibinfo {author} {\bibfnamefont {S.}~\bibnamefont {Burchesky}}, \bibinfo {author} {\bibfnamefont {W.}~\bibnamefont {Ketterle}}, \ and\ \bibinfo {author} {\bibfnamefont {J.~M.}\ \bibnamefont {Doyle}},\ }\href@noop {} {\bibfield  {journal} {\bibinfo  {journal} {Physical review letters}\ }\textbf {\bibinfo {volume} {121}},\ \bibinfo {pages} {083201} (\bibinfo {year} {2018})}\BibitemShut {NoStop}%
\bibitem [{\citenamefont {Lu}\ \emph {et~al.}(2024)\citenamefont {Lu}, \citenamefont {Li}, \citenamefont {Holland},\ and\ \citenamefont {Cheuk}}]{lu2024raman}%
  \BibitemOpen
  \bibfield  {author} {\bibinfo {author} {\bibfnamefont {Y.}~\bibnamefont {Lu}}, \bibinfo {author} {\bibfnamefont {S.~J.}\ \bibnamefont {Li}}, \bibinfo {author} {\bibfnamefont {C.~M.}\ \bibnamefont {Holland}}, \ and\ \bibinfo {author} {\bibfnamefont {L.~W.}\ \bibnamefont {Cheuk}},\ }\href@noop {} {\bibfield  {journal} {\bibinfo  {journal} {Nature Physics}\ }\textbf {\bibinfo {volume} {20}},\ \bibinfo {pages} {389} (\bibinfo {year} {2024})}\BibitemShut {NoStop}%
\bibitem [{\citenamefont {Caldwell}\ and\ \citenamefont {Tarbutt}(2020)}]{caldwell2020sideband}%
  \BibitemOpen
  \bibfield  {author} {\bibinfo {author} {\bibfnamefont {L.}~\bibnamefont {Caldwell}}\ and\ \bibinfo {author} {\bibfnamefont {M.}~\bibnamefont {Tarbutt}},\ }\href@noop {} {\bibfield  {journal} {\bibinfo  {journal} {Physical Review Research}\ }\textbf {\bibinfo {volume} {2}},\ \bibinfo {pages} {013251} (\bibinfo {year} {2020})}\BibitemShut {NoStop}%
\bibitem [{\citenamefont {Bao}\ \emph {et~al.}(2024)\citenamefont {Bao}, \citenamefont {Yu}, \citenamefont {You}, \citenamefont {Anderegg}, \citenamefont {Chae}, \citenamefont {Ketterle}, \citenamefont {Ni},\ and\ \citenamefont {Doyle}}]{bao2024raman}%
  \BibitemOpen
  \bibfield  {author} {\bibinfo {author} {\bibfnamefont {Y.}~\bibnamefont {Bao}}, \bibinfo {author} {\bibfnamefont {S.~S.}\ \bibnamefont {Yu}}, \bibinfo {author} {\bibfnamefont {J.}~\bibnamefont {You}}, \bibinfo {author} {\bibfnamefont {L.}~\bibnamefont {Anderegg}}, \bibinfo {author} {\bibfnamefont {E.}~\bibnamefont {Chae}}, \bibinfo {author} {\bibfnamefont {W.}~\bibnamefont {Ketterle}}, \bibinfo {author} {\bibfnamefont {K.-K.}\ \bibnamefont {Ni}}, \ and\ \bibinfo {author} {\bibfnamefont {J.~M.}\ \bibnamefont {Doyle}},\ }\href@noop {} {\bibfield  {journal} {\bibinfo  {journal} {Physical Review X}\ }\textbf {\bibinfo {volume} {14}},\ \bibinfo {pages} {031002} (\bibinfo {year} {2024})}\BibitemShut {NoStop}%
\bibitem [{SM()}]{SM}%
  \BibitemOpen
  \href@noop {} {}\bibinfo {note} {See Supplemental Material for additional details}\BibitemShut {NoStop}%
\bibitem [{\citenamefont {Emperauger}\ \emph {et~al.}(2025)\citenamefont {Emperauger}, \citenamefont {Qiao}, \citenamefont {Bornet}, \citenamefont {Chen}, \citenamefont {Martin}, \citenamefont {Chew}, \citenamefont {G{\'e}ly}, \citenamefont {Klein}, \citenamefont {Barredo}, \citenamefont {Browaeys} \emph {et~al.}}]{emperauger2025benchmarking}%
  \BibitemOpen
  \bibfield  {author} {\bibinfo {author} {\bibfnamefont {G.}~\bibnamefont {Emperauger}}, \bibinfo {author} {\bibfnamefont {M.}~\bibnamefont {Qiao}}, \bibinfo {author} {\bibfnamefont {G.}~\bibnamefont {Bornet}}, \bibinfo {author} {\bibfnamefont {C.}~\bibnamefont {Chen}}, \bibinfo {author} {\bibfnamefont {R.}~\bibnamefont {Martin}}, \bibinfo {author} {\bibfnamefont {Y.~T.}\ \bibnamefont {Chew}}, \bibinfo {author} {\bibfnamefont {B.}~\bibnamefont {G{\'e}ly}}, \bibinfo {author} {\bibfnamefont {L.}~\bibnamefont {Klein}}, \bibinfo {author} {\bibfnamefont {D.}~\bibnamefont {Barredo}}, \bibinfo {author} {\bibfnamefont {A.}~\bibnamefont {Browaeys}},  \emph {et~al.},\ }\href@noop {} {\bibfield  {journal} {\bibinfo  {journal} {Physical Review A}\ }\textbf {\bibinfo {volume} {111}},\ \bibinfo {pages} {062806} (\bibinfo {year} {2025})}\BibitemShut {NoStop}%
\bibitem [{\citenamefont {Perlin}\ \emph {et~al.}(2020)\citenamefont {Perlin}, \citenamefont {Qu},\ and\ \citenamefont {Rey}}]{perlin2020spin}%
  \BibitemOpen
  \bibfield  {author} {\bibinfo {author} {\bibfnamefont {M.~A.}\ \bibnamefont {Perlin}}, \bibinfo {author} {\bibfnamefont {C.}~\bibnamefont {Qu}}, \ and\ \bibinfo {author} {\bibfnamefont {A.~M.}\ \bibnamefont {Rey}},\ }\href@noop {} {\bibfield  {journal} {\bibinfo  {journal} {Physical Review Letters}\ }\textbf {\bibinfo {volume} {125}},\ \bibinfo {pages} {223401} (\bibinfo {year} {2020})}\BibitemShut {NoStop}%
\bibitem [{\citenamefont {Eckner}\ \emph {et~al.}(2023)\citenamefont {Eckner}, \citenamefont {Darkwah~Oppong}, \citenamefont {Cao}, \citenamefont {Young}, \citenamefont {Milner}, \citenamefont {Robinson}, \citenamefont {Ye},\ and\ \citenamefont {Kaufman}}]{eckner2023realizing}%
  \BibitemOpen
  \bibfield  {author} {\bibinfo {author} {\bibfnamefont {W.~J.}\ \bibnamefont {Eckner}}, \bibinfo {author} {\bibfnamefont {N.}~\bibnamefont {Darkwah~Oppong}}, \bibinfo {author} {\bibfnamefont {A.}~\bibnamefont {Cao}}, \bibinfo {author} {\bibfnamefont {A.~W.}\ \bibnamefont {Young}}, \bibinfo {author} {\bibfnamefont {W.~R.}\ \bibnamefont {Milner}}, \bibinfo {author} {\bibfnamefont {J.~M.}\ \bibnamefont {Robinson}}, \bibinfo {author} {\bibfnamefont {J.}~\bibnamefont {Ye}}, \ and\ \bibinfo {author} {\bibfnamefont {A.~M.}\ \bibnamefont {Kaufman}},\ }\href@noop {} {\bibfield  {journal} {\bibinfo  {journal} {Nature}\ }\textbf {\bibinfo {volume} {621}},\ \bibinfo {pages} {734} (\bibinfo {year} {2023})}\BibitemShut {NoStop}%
\bibitem [{\citenamefont {Hines}\ \emph {et~al.}(2023)\citenamefont {Hines}, \citenamefont {Rajagopal}, \citenamefont {Moreau}, \citenamefont {Wahrman}, \citenamefont {Lewis}, \citenamefont {Markovi{\'c}},\ and\ \citenamefont {Schleier-Smith}}]{hines2023spin}%
  \BibitemOpen
  \bibfield  {author} {\bibinfo {author} {\bibfnamefont {J.~A.}\ \bibnamefont {Hines}}, \bibinfo {author} {\bibfnamefont {S.~V.}\ \bibnamefont {Rajagopal}}, \bibinfo {author} {\bibfnamefont {G.~L.}\ \bibnamefont {Moreau}}, \bibinfo {author} {\bibfnamefont {M.~D.}\ \bibnamefont {Wahrman}}, \bibinfo {author} {\bibfnamefont {N.~A.}\ \bibnamefont {Lewis}}, \bibinfo {author} {\bibfnamefont {O.}~\bibnamefont {Markovi{\'c}}}, \ and\ \bibinfo {author} {\bibfnamefont {M.}~\bibnamefont {Schleier-Smith}},\ }\href@noop {} {\bibfield  {journal} {\bibinfo  {journal} {Physical Review Letters}\ }\textbf {\bibinfo {volume} {131}},\ \bibinfo {pages} {063401} (\bibinfo {year} {2023})}\BibitemShut {NoStop}%
\bibitem [{\citenamefont {Block}\ \emph {et~al.}(2024)\citenamefont {Block}, \citenamefont {Ye}, \citenamefont {Roberts}, \citenamefont {Chern}, \citenamefont {Wu}, \citenamefont {Wang}, \citenamefont {Pollet}, \citenamefont {Davis}, \citenamefont {Halperin},\ and\ \citenamefont {Yao}}]{block2024scalable}%
  \BibitemOpen
  \bibfield  {author} {\bibinfo {author} {\bibfnamefont {M.}~\bibnamefont {Block}}, \bibinfo {author} {\bibfnamefont {B.}~\bibnamefont {Ye}}, \bibinfo {author} {\bibfnamefont {B.}~\bibnamefont {Roberts}}, \bibinfo {author} {\bibfnamefont {S.}~\bibnamefont {Chern}}, \bibinfo {author} {\bibfnamefont {W.}~\bibnamefont {Wu}}, \bibinfo {author} {\bibfnamefont {Z.}~\bibnamefont {Wang}}, \bibinfo {author} {\bibfnamefont {L.}~\bibnamefont {Pollet}}, \bibinfo {author} {\bibfnamefont {E.~J.}\ \bibnamefont {Davis}}, \bibinfo {author} {\bibfnamefont {B.~I.}\ \bibnamefont {Halperin}}, \ and\ \bibinfo {author} {\bibfnamefont {N.~Y.}\ \bibnamefont {Yao}},\ }\href@noop {} {\bibfield  {journal} {\bibinfo  {journal} {Nature Physics}\ }\textbf {\bibinfo {volume} {20}},\ \bibinfo {pages} {1575} (\bibinfo {year} {2024})}\BibitemShut {NoStop}%
\bibitem [{\citenamefont {Douglas}\ \emph {et~al.}(2024)\citenamefont {Douglas}, \citenamefont {Kaxiras}, \citenamefont {Su}, \citenamefont {Szurek}, \citenamefont {Singh}, \citenamefont {Markovi{\'c}},\ and\ \citenamefont {Greiner}}]{douglas2024spin}%
  \BibitemOpen
  \bibfield  {author} {\bibinfo {author} {\bibfnamefont {A.}~\bibnamefont {Douglas}}, \bibinfo {author} {\bibfnamefont {V.}~\bibnamefont {Kaxiras}}, \bibinfo {author} {\bibfnamefont {L.}~\bibnamefont {Su}}, \bibinfo {author} {\bibfnamefont {M.}~\bibnamefont {Szurek}}, \bibinfo {author} {\bibfnamefont {V.}~\bibnamefont {Singh}}, \bibinfo {author} {\bibfnamefont {O.}~\bibnamefont {Markovi{\'c}}}, \ and\ \bibinfo {author} {\bibfnamefont {M.}~\bibnamefont {Greiner}},\ }\href@noop {} {\bibfield  {journal} {\bibinfo  {journal} {arXiv preprint arXiv:2411.07219}\ } (\bibinfo {year} {2024})}\BibitemShut {NoStop}%
\bibitem [{\citenamefont {Wineland}\ \emph {et~al.}(1994)\citenamefont {Wineland}, \citenamefont {Bollinger}, \citenamefont {Itano},\ and\ \citenamefont {Heinzen}}]{wineland1994squeezed}%
  \BibitemOpen
  \bibfield  {author} {\bibinfo {author} {\bibfnamefont {D.~J.}\ \bibnamefont {Wineland}}, \bibinfo {author} {\bibfnamefont {J.~J.}\ \bibnamefont {Bollinger}}, \bibinfo {author} {\bibfnamefont {W.~M.}\ \bibnamefont {Itano}}, \ and\ \bibinfo {author} {\bibfnamefont {D.~J.}\ \bibnamefont {Heinzen}},\ }\href@noop {} {\bibfield  {journal} {\bibinfo  {journal} {Physical Review A}\ }\textbf {\bibinfo {volume} {50}},\ \bibinfo {pages} {67} (\bibinfo {year} {1994})}\BibitemShut {NoStop}%
\bibitem [{\citenamefont {Wellnitz}\ \emph {et~al.}(2024)\citenamefont {Wellnitz}, \citenamefont {Mamaev}, \citenamefont {Bilitewski},\ and\ \citenamefont {Rey}}]{wellnitz2024spin}%
  \BibitemOpen
  \bibfield  {author} {\bibinfo {author} {\bibfnamefont {D.}~\bibnamefont {Wellnitz}}, \bibinfo {author} {\bibfnamefont {M.}~\bibnamefont {Mamaev}}, \bibinfo {author} {\bibfnamefont {T.}~\bibnamefont {Bilitewski}}, \ and\ \bibinfo {author} {\bibfnamefont {A.~M.}\ \bibnamefont {Rey}},\ }\href@noop {} {\bibfield  {journal} {\bibinfo  {journal} {Physical Review Research}\ }\textbf {\bibinfo {volume} {6}},\ \bibinfo {pages} {L012025} (\bibinfo {year} {2024})}\BibitemShut {NoStop}%
\bibitem [{\citenamefont {Schachenmayer}\ \emph {et~al.}(2015)\citenamefont {Schachenmayer}, \citenamefont {Pikovski},\ and\ \citenamefont {Rey}}]{schachenmayer2015many}%
  \BibitemOpen
  \bibfield  {author} {\bibinfo {author} {\bibfnamefont {J.}~\bibnamefont {Schachenmayer}}, \bibinfo {author} {\bibfnamefont {A.}~\bibnamefont {Pikovski}}, \ and\ \bibinfo {author} {\bibfnamefont {A.~M.}\ \bibnamefont {Rey}},\ }\href@noop {} {\bibfield  {journal} {\bibinfo  {journal} {Physical Review X}\ }\textbf {\bibinfo {volume} {5}},\ \bibinfo {pages} {011022} (\bibinfo {year} {2015})}\BibitemShut {NoStop}%
\bibitem [{\citenamefont {Borish}\ \emph {et~al.}(2020)\citenamefont {Borish}, \citenamefont {Markovi{\'c}}, \citenamefont {Hines}, \citenamefont {Rajagopal},\ and\ \citenamefont {Schleier-Smith}}]{borish2020transverse}%
  \BibitemOpen
  \bibfield  {author} {\bibinfo {author} {\bibfnamefont {V.}~\bibnamefont {Borish}}, \bibinfo {author} {\bibfnamefont {O.}~\bibnamefont {Markovi{\'c}}}, \bibinfo {author} {\bibfnamefont {J.~A.}\ \bibnamefont {Hines}}, \bibinfo {author} {\bibfnamefont {S.~V.}\ \bibnamefont {Rajagopal}}, \ and\ \bibinfo {author} {\bibfnamefont {M.}~\bibnamefont {Schleier-Smith}},\ }\href@noop {} {\bibfield  {journal} {\bibinfo  {journal} {Physical review letters}\ }\textbf {\bibinfo {volume} {124}},\ \bibinfo {pages} {063601} (\bibinfo {year} {2020})}\BibitemShut {NoStop}%
\bibitem [{\citenamefont {Schine}\ \emph {et~al.}(2022)\citenamefont {Schine}, \citenamefont {Young}, \citenamefont {Eckner}, \citenamefont {Martin},\ and\ \citenamefont {Kaufman}}]{schine2022long}%
  \BibitemOpen
  \bibfield  {author} {\bibinfo {author} {\bibfnamefont {N.}~\bibnamefont {Schine}}, \bibinfo {author} {\bibfnamefont {A.~W.}\ \bibnamefont {Young}}, \bibinfo {author} {\bibfnamefont {W.~J.}\ \bibnamefont {Eckner}}, \bibinfo {author} {\bibfnamefont {M.~J.}\ \bibnamefont {Martin}}, \ and\ \bibinfo {author} {\bibfnamefont {A.~M.}\ \bibnamefont {Kaufman}},\ }\href@noop {} {\bibfield  {journal} {\bibinfo  {journal} {Nature Physics}\ }\textbf {\bibinfo {volume} {18}},\ \bibinfo {pages} {1067} (\bibinfo {year} {2022})}\BibitemShut {NoStop}%
\bibitem [{\citenamefont {Ma}\ \emph {et~al.}(2023)\citenamefont {Ma}, \citenamefont {Liu}, \citenamefont {Peng}, \citenamefont {Zhang}, \citenamefont {Jandura}, \citenamefont {Claes}, \citenamefont {Burgers}, \citenamefont {Pupillo}, \citenamefont {Puri},\ and\ \citenamefont {Thompson}}]{ma2023high}%
  \BibitemOpen
  \bibfield  {author} {\bibinfo {author} {\bibfnamefont {S.}~\bibnamefont {Ma}}, \bibinfo {author} {\bibfnamefont {G.}~\bibnamefont {Liu}}, \bibinfo {author} {\bibfnamefont {P.}~\bibnamefont {Peng}}, \bibinfo {author} {\bibfnamefont {B.}~\bibnamefont {Zhang}}, \bibinfo {author} {\bibfnamefont {S.}~\bibnamefont {Jandura}}, \bibinfo {author} {\bibfnamefont {J.}~\bibnamefont {Claes}}, \bibinfo {author} {\bibfnamefont {A.~P.}\ \bibnamefont {Burgers}}, \bibinfo {author} {\bibfnamefont {G.}~\bibnamefont {Pupillo}}, \bibinfo {author} {\bibfnamefont {S.}~\bibnamefont {Puri}}, \ and\ \bibinfo {author} {\bibfnamefont {J.~D.}\ \bibnamefont {Thompson}},\ }\href@noop {} {\bibfield  {journal} {\bibinfo  {journal} {Nature}\ }\textbf {\bibinfo {volume} {622}},\ \bibinfo {pages} {279} (\bibinfo {year} {2023})}\BibitemShut {NoStop}%
\bibitem [{\citenamefont {Scholl}\ \emph {et~al.}(2023)\citenamefont {Scholl}, \citenamefont {Shaw}, \citenamefont {Tsai}, \citenamefont {Finkelstein}, \citenamefont {Choi},\ and\ \citenamefont {Endres}}]{scholl2023erasure}%
  \BibitemOpen
  \bibfield  {author} {\bibinfo {author} {\bibfnamefont {P.}~\bibnamefont {Scholl}}, \bibinfo {author} {\bibfnamefont {A.~L.}\ \bibnamefont {Shaw}}, \bibinfo {author} {\bibfnamefont {R.~B.-S.}\ \bibnamefont {Tsai}}, \bibinfo {author} {\bibfnamefont {R.}~\bibnamefont {Finkelstein}}, \bibinfo {author} {\bibfnamefont {J.}~\bibnamefont {Choi}}, \ and\ \bibinfo {author} {\bibfnamefont {M.}~\bibnamefont {Endres}},\ }\href@noop {} {\bibfield  {journal} {\bibinfo  {journal} {Nature}\ }\textbf {\bibinfo {volume} {622}},\ \bibinfo {pages} {273} (\bibinfo {year} {2023})}\BibitemShut {NoStop}%
\end{thebibliography}%


\providecommand{\noopsort}[1]{}\providecommand{\singleletter}[1]{#1}%
\begin{thebibliography}{12}%
\makeatletter
\providecommand \@ifxundefined [1]{%
 \@ifx{#1\undefined}
}%
\providecommand \@ifnum [1]{%
 \ifnum #1\expandafter \@firstoftwo
 \else \expandafter \@secondoftwo
 \fi
}%
\providecommand \@ifx [1]{%
 \ifx #1\expandafter \@firstoftwo
 \else \expandafter \@secondoftwo
 \fi
}%
\providecommand \natexlab [1]{#1}%
\providecommand \enquote  [1]{``#1''}%
\providecommand \bibnamefont  [1]{#1}%
\providecommand \bibfnamefont [1]{#1}%
\providecommand \citenamefont [1]{#1}%
\providecommand \href@noop [0]{\@secondoftwo}%
\providecommand \href [0]{\begingroup \@sanitize@url \@href}%
\providecommand \@href[1]{\@@startlink{#1}\@@href}%
\providecommand \@@href[1]{\endgroup#1\@@endlink}%
\providecommand \@sanitize@url [0]{\catcode `\\12\catcode `\$12\catcode `\&12\catcode `\#12\catcode `\^12\catcode `\_12\catcode `\%12\relax}%
\providecommand \@@startlink[1]{}%
\providecommand \@@endlink[0]{}%
\providecommand \url  [0]{\begingroup\@sanitize@url \@url }%
\providecommand \@url [1]{\endgroup\@href {#1}{\urlprefix }}%
\providecommand \urlprefix  [0]{URL }%
\providecommand \Eprint [0]{\href }%
\providecommand \doibase [0]{http://dx.doi.org/}%
\providecommand \selectlanguage [0]{\@gobble}%
\providecommand \bibinfo  [0]{\@secondoftwo}%
\providecommand \bibfield  [0]{\@secondoftwo}%
\providecommand \translation [1]{[#1]}%
\providecommand \BibitemOpen [0]{}%
\providecommand \bibitemStop [0]{}%
\providecommand \bibitemNoStop [0]{.\EOS\space}%
\providecommand \EOS [0]{\spacefactor3000\relax}%
\providecommand \BibitemShut  [1]{\csname bibitem#1\endcsname}%
\let\auto@bib@innerbib\@empty
%</preamble>
\bibitem [{\citenamefont {Ni}\ \emph {et~al.}(2018)\citenamefont {Ni}, \citenamefont {Rosenband},\ and\ \citenamefont {Grimes}}]{ni2018dipolar}%
  \BibitemOpen
  \bibfield  {author} {\bibinfo {author} {\bibfnamefont {K.-K.}\ \bibnamefont {Ni}}, \bibinfo {author} {\bibfnamefont {T.}~\bibnamefont {Rosenband}}, \ and\ \bibinfo {author} {\bibfnamefont {D.~D.}\ \bibnamefont {Grimes}},\ }\href@noop {} {\bibfield  {journal} {\bibinfo  {journal} {Chemical science}\ }\textbf {\bibinfo {volume} {9}},\ \bibinfo {pages} {6830} (\bibinfo {year} {2018})}\BibitemShut {NoStop}%
\bibitem [{\citenamefont {Bao}\ \emph {et~al.}(2023)\citenamefont {Bao}, \citenamefont {Yu}, \citenamefont {Anderegg}, \citenamefont {Chae}, \citenamefont {Ketterle}, \citenamefont {Ni},\ and\ \citenamefont {Doyle}}]{bao2023dipolar}%
  \BibitemOpen
  \bibfield  {author} {\bibinfo {author} {\bibfnamefont {Y.}~\bibnamefont {Bao}}, \bibinfo {author} {\bibfnamefont {S.~S.}\ \bibnamefont {Yu}}, \bibinfo {author} {\bibfnamefont {L.}~\bibnamefont {Anderegg}}, \bibinfo {author} {\bibfnamefont {E.}~\bibnamefont {Chae}}, \bibinfo {author} {\bibfnamefont {W.}~\bibnamefont {Ketterle}}, \bibinfo {author} {\bibfnamefont {K.-K.}\ \bibnamefont {Ni}}, \ and\ \bibinfo {author} {\bibfnamefont {J.~M.}\ \bibnamefont {Doyle}},\ }\href@noop {} {\bibfield  {journal} {\bibinfo  {journal} {Science}\ }\textbf {\bibinfo {volume} {382}},\ \bibinfo {pages} {1138} (\bibinfo {year} {2023})}\BibitemShut {NoStop}%
\bibitem [{\citenamefont {Holland}\ \emph {et~al.}(2023)\citenamefont {Holland}, \citenamefont {Lu},\ and\ \citenamefont {Cheuk}}]{holland2023demand}%
  \BibitemOpen
  \bibfield  {author} {\bibinfo {author} {\bibfnamefont {C.~M.}\ \bibnamefont {Holland}}, \bibinfo {author} {\bibfnamefont {Y.}~\bibnamefont {Lu}}, \ and\ \bibinfo {author} {\bibfnamefont {L.~W.}\ \bibnamefont {Cheuk}},\ }\href@noop {} {\bibfield  {journal} {\bibinfo  {journal} {Science}\ }\textbf {\bibinfo {volume} {382}},\ \bibinfo {pages} {1143} (\bibinfo {year} {2023})}\BibitemShut {NoStop}%
\bibitem [{\citenamefont {Picard}\ \emph {et~al.}(2025)\citenamefont {Picard}, \citenamefont {Park}, \citenamefont {Patenotte}, \citenamefont {Gebretsadkan}, \citenamefont {Wellnitz}, \citenamefont {Rey},\ and\ \citenamefont {Ni}}]{picard2025entanglement}%
  \BibitemOpen
  \bibfield  {author} {\bibinfo {author} {\bibfnamefont {L.~R.}\ \bibnamefont {Picard}}, \bibinfo {author} {\bibfnamefont {A.~J.}\ \bibnamefont {Park}}, \bibinfo {author} {\bibfnamefont {G.~E.}\ \bibnamefont {Patenotte}}, \bibinfo {author} {\bibfnamefont {S.}~\bibnamefont {Gebretsadkan}}, \bibinfo {author} {\bibfnamefont {D.}~\bibnamefont {Wellnitz}}, \bibinfo {author} {\bibfnamefont {A.~M.}\ \bibnamefont {Rey}}, \ and\ \bibinfo {author} {\bibfnamefont {K.-K.}\ \bibnamefont {Ni}},\ }\href@noop {} {\bibfield  {journal} {\bibinfo  {journal} {Nature}\ }\textbf {\bibinfo {volume} {637}},\ \bibinfo {pages} {821} (\bibinfo {year} {2025})}\BibitemShut {NoStop}%
\bibitem [{\citenamefont {Ruttley}\ \emph {et~al.}(2025)\citenamefont {Ruttley}, \citenamefont {Hepworth}, \citenamefont {Guttridge},\ and\ \citenamefont {Cornish}}]{ruttley2025long}%
  \BibitemOpen
  \bibfield  {author} {\bibinfo {author} {\bibfnamefont {D.~K.}\ \bibnamefont {Ruttley}}, \bibinfo {author} {\bibfnamefont {T.~R.}\ \bibnamefont {Hepworth}}, \bibinfo {author} {\bibfnamefont {A.}~\bibnamefont {Guttridge}}, \ and\ \bibinfo {author} {\bibfnamefont {S.~L.}\ \bibnamefont {Cornish}},\ }\href@noop {} {\bibfield  {journal} {\bibinfo  {journal} {Nature}\ ,\ \bibinfo {pages} {1}} (\bibinfo {year} {2025})}\BibitemShut {NoStop}%
\bibitem [{\citenamefont {Yan}\ \emph {et~al.}(2013)\citenamefont {Yan}, \citenamefont {Moses}, \citenamefont {Gadway}, \citenamefont {Covey}, \citenamefont {Hazzard}, \citenamefont {Rey}, \citenamefont {Jin},\ and\ \citenamefont {Ye}}]{yan2013observation}%
  \BibitemOpen
  \bibfield  {author} {\bibinfo {author} {\bibfnamefont {B.}~\bibnamefont {Yan}}, \bibinfo {author} {\bibfnamefont {S.~A.}\ \bibnamefont {Moses}}, \bibinfo {author} {\bibfnamefont {B.}~\bibnamefont {Gadway}}, \bibinfo {author} {\bibfnamefont {J.~P.}\ \bibnamefont {Covey}}, \bibinfo {author} {\bibfnamefont {K.~R.}\ \bibnamefont {Hazzard}}, \bibinfo {author} {\bibfnamefont {A.~M.}\ \bibnamefont {Rey}}, \bibinfo {author} {\bibfnamefont {D.~S.}\ \bibnamefont {Jin}}, \ and\ \bibinfo {author} {\bibfnamefont {J.}~\bibnamefont {Ye}},\ }\href@noop {} {\bibfield  {journal} {\bibinfo  {journal} {Nature}\ }\textbf {\bibinfo {volume} {501}},\ \bibinfo {pages} {521} (\bibinfo {year} {2013})}\BibitemShut {NoStop}%
\bibitem [{\citenamefont {Christakis}\ \emph {et~al.}(2023)\citenamefont {Christakis}, \citenamefont {Rosenberg}, \citenamefont {Raj}, \citenamefont {Chi}, \citenamefont {Morningstar}, \citenamefont {Huse}, \citenamefont {Yan},\ and\ \citenamefont {Bakr}}]{christakis2023probing}%
  \BibitemOpen
  \bibfield  {author} {\bibinfo {author} {\bibfnamefont {L.}~\bibnamefont {Christakis}}, \bibinfo {author} {\bibfnamefont {J.~S.}\ \bibnamefont {Rosenberg}}, \bibinfo {author} {\bibfnamefont {R.}~\bibnamefont {Raj}}, \bibinfo {author} {\bibfnamefont {S.}~\bibnamefont {Chi}}, \bibinfo {author} {\bibfnamefont {A.}~\bibnamefont {Morningstar}}, \bibinfo {author} {\bibfnamefont {D.~A.}\ \bibnamefont {Huse}}, \bibinfo {author} {\bibfnamefont {Z.~Z.}\ \bibnamefont {Yan}}, \ and\ \bibinfo {author} {\bibfnamefont {W.~S.}\ \bibnamefont {Bakr}},\ }\href@noop {} {\bibfield  {journal} {\bibinfo  {journal} {Nature}\ }\textbf {\bibinfo {volume} {614}},\ \bibinfo {pages} {64} (\bibinfo {year} {2023})}\BibitemShut {NoStop}%
\bibitem [{\citenamefont {Evered}\ \emph {et~al.}(2023)\citenamefont {Evered}, \citenamefont {Bluvstein}, \citenamefont {Kalinowski}, \citenamefont {Ebadi}, \citenamefont {Manovitz}, \citenamefont {Zhou}, \citenamefont {Li}, \citenamefont {Geim}, \citenamefont {Wang}, \citenamefont {Maskara} \emph {et~al.}}]{evered2023high}%
  \BibitemOpen
  \bibfield  {author} {\bibinfo {author} {\bibfnamefont {S.~J.}\ \bibnamefont {Evered}}, \bibinfo {author} {\bibfnamefont {D.}~\bibnamefont {Bluvstein}}, \bibinfo {author} {\bibfnamefont {M.}~\bibnamefont {Kalinowski}}, \bibinfo {author} {\bibfnamefont {S.}~\bibnamefont {Ebadi}}, \bibinfo {author} {\bibfnamefont {T.}~\bibnamefont {Manovitz}}, \bibinfo {author} {\bibfnamefont {H.}~\bibnamefont {Zhou}}, \bibinfo {author} {\bibfnamefont {S.~H.}\ \bibnamefont {Li}}, \bibinfo {author} {\bibfnamefont {A.~A.}\ \bibnamefont {Geim}}, \bibinfo {author} {\bibfnamefont {T.~T.}\ \bibnamefont {Wang}}, \bibinfo {author} {\bibfnamefont {N.}~\bibnamefont {Maskara}},  \emph {et~al.},\ }\href@noop {} {\bibfield  {journal} {\bibinfo  {journal} {Nature}\ }\textbf {\bibinfo {volume} {622}},\ \bibinfo {pages} {268} (\bibinfo {year} {2023})}\BibitemShut {NoStop}%
\bibitem [{\citenamefont {Block}\ \emph {et~al.}(2024)\citenamefont {Block}, \citenamefont {Ye}, \citenamefont {Roberts}, \citenamefont {Chern}, \citenamefont {Wu}, \citenamefont {Wang}, \citenamefont {Pollet}, \citenamefont {Davis}, \citenamefont {Halperin},\ and\ \citenamefont {Yao}}]{block2024scalable}%
  \BibitemOpen
  \bibfield  {author} {\bibinfo {author} {\bibfnamefont {M.}~\bibnamefont {Block}}, \bibinfo {author} {\bibfnamefont {B.}~\bibnamefont {Ye}}, \bibinfo {author} {\bibfnamefont {B.}~\bibnamefont {Roberts}}, \bibinfo {author} {\bibfnamefont {S.}~\bibnamefont {Chern}}, \bibinfo {author} {\bibfnamefont {W.}~\bibnamefont {Wu}}, \bibinfo {author} {\bibfnamefont {Z.}~\bibnamefont {Wang}}, \bibinfo {author} {\bibfnamefont {L.}~\bibnamefont {Pollet}}, \bibinfo {author} {\bibfnamefont {E.~J.}\ \bibnamefont {Davis}}, \bibinfo {author} {\bibfnamefont {B.~I.}\ \bibnamefont {Halperin}}, \ and\ \bibinfo {author} {\bibfnamefont {N.~Y.}\ \bibnamefont {Yao}},\ }\href@noop {} {\bibfield  {journal} {\bibinfo  {journal} {Nature Physics}\ }\textbf {\bibinfo {volume} {20}},\ \bibinfo {pages} {1575} (\bibinfo {year} {2024})}\BibitemShut {NoStop}%
\bibitem [{\citenamefont {Bentsen}\ \emph {et~al.}(2019)\citenamefont {Bentsen}, \citenamefont {Hashizume}, \citenamefont {Buyskikh}, \citenamefont {Davis}, \citenamefont {Daley}, \citenamefont {Gubser},\ and\ \citenamefont {Schleier-Smith}}]{bentsen2019treelike}%
  \BibitemOpen
  \bibfield  {author} {\bibinfo {author} {\bibfnamefont {G.}~\bibnamefont {Bentsen}}, \bibinfo {author} {\bibfnamefont {T.}~\bibnamefont {Hashizume}}, \bibinfo {author} {\bibfnamefont {A.~S.}\ \bibnamefont {Buyskikh}}, \bibinfo {author} {\bibfnamefont {E.~J.}\ \bibnamefont {Davis}}, \bibinfo {author} {\bibfnamefont {A.~J.}\ \bibnamefont {Daley}}, \bibinfo {author} {\bibfnamefont {S.~S.}\ \bibnamefont {Gubser}}, \ and\ \bibinfo {author} {\bibfnamefont {M.}~\bibnamefont {Schleier-Smith}},\ }\href@noop {} {\bibfield  {journal} {\bibinfo  {journal} {Physical review letters}\ }\textbf {\bibinfo {volume} {123}},\ \bibinfo {pages} {130601} (\bibinfo {year} {2019})}\BibitemShut {NoStop}%
\bibitem [{\citenamefont {Hashizume}\ \emph {et~al.}(2021)\citenamefont {Hashizume}, \citenamefont {Bentsen}, \citenamefont {Weber},\ and\ \citenamefont {Daley}}]{hashizume2021deterministic}%
  \BibitemOpen
  \bibfield  {author} {\bibinfo {author} {\bibfnamefont {T.}~\bibnamefont {Hashizume}}, \bibinfo {author} {\bibfnamefont {G.~S.}\ \bibnamefont {Bentsen}}, \bibinfo {author} {\bibfnamefont {S.}~\bibnamefont {Weber}}, \ and\ \bibinfo {author} {\bibfnamefont {A.~J.}\ \bibnamefont {Daley}},\ }\href@noop {} {\bibfield  {journal} {\bibinfo  {journal} {Physical Review Letters}\ }\textbf {\bibinfo {volume} {126}},\ \bibinfo {pages} {200603} (\bibinfo {year} {2021})}\BibitemShut {NoStop}%
\bibitem [{\citenamefont {Lin}\ \emph {et~al.}(2024)\citenamefont {Lin}, \citenamefont {Zhong}, \citenamefont {Li}, \citenamefont {Zhao}, \citenamefont {Zheng}, \citenamefont {Hu}, \citenamefont {Wu}, \citenamefont {Wu}, \citenamefont {Ma}, \citenamefont {Gao} \emph {et~al.}}]{lin2024ai}%
  \BibitemOpen
  \bibfield  {author} {\bibinfo {author} {\bibfnamefont {R.}~\bibnamefont {Lin}}, \bibinfo {author} {\bibfnamefont {H.-S.}\ \bibnamefont {Zhong}}, \bibinfo {author} {\bibfnamefont {Y.}~\bibnamefont {Li}}, \bibinfo {author} {\bibfnamefont {Z.-R.}\ \bibnamefont {Zhao}}, \bibinfo {author} {\bibfnamefont {L.-T.}\ \bibnamefont {Zheng}}, \bibinfo {author} {\bibfnamefont {T.-R.}\ \bibnamefont {Hu}}, \bibinfo {author} {\bibfnamefont {H.-M.}\ \bibnamefont {Wu}}, \bibinfo {author} {\bibfnamefont {Z.}~\bibnamefont {Wu}}, \bibinfo {author} {\bibfnamefont {W.-J.}\ \bibnamefont {Ma}}, \bibinfo {author} {\bibfnamefont {Y.}~\bibnamefont {Gao}},  \emph {et~al.},\ }\href@noop {} {\bibfield  {journal} {\bibinfo  {journal} {arXiv preprint arXiv:2412.14647}\ } (\bibinfo {year} {2024})}\BibitemShut {NoStop}%
\end{thebibliography}%
% }{}
\end{document}
\putbib
\end{bibunit}
\end{document}